\documentclass[12pt,journal,a4paper,onecolumn]{IEEEtran}

\usepackage{amsmath,epsfig,amssymb,verbatim,amsopn,cite,subfigure,multirow}
\usepackage{balance}
\usepackage[usenames,dvipsnames]{color}
\usepackage{url}
\usepackage{amsfonts}
\usepackage{amssymb}
\usepackage{epsfig}
\usepackage{slashbox}
\usepackage{bm}
\usepackage{epsfig}
\usepackage{amsmath,amssymb, amstext}
\usepackage{graphicx}
\usepackage{epstopdf}
\newtheorem{theo}{Theorem}
\newtheorem{lem}{Lemma}
\newtheorem{remk}{Remark}
\newtheorem{defin}{Definition}

\usepackage[T1]{fontenc}
\usepackage[table]{xcolor}
\usepackage{psfrag}
\usepackage{setspace}
\usepackage[process=auto]{pstool}
\usepackage{etoolbox}
\allowdisplaybreaks

\newtoggle{OneColumn}

\toggletrue{OneColumn}

\iftoggle{OneColumn}{%
  \doublespacing
}{%
}

\newcommand{\Equal}{\hspace{-0.7mm}=\hspace{-0.7mm}}
\newcommand{\Add}{\hspace{-0.7mm}+\hspace{-0.7mm}}
\newcommand{\Minus}{\hspace{-0.7mm}-\hspace{-0.7mm}}

\newcommand{\Great}{\hspace{-0.7mm}>\hspace{-0.7mm}}

\EndPreamble
\begin{document}

\title{On the Design of Fast Convergent LDPC Codes: An Optimization Approach }


\author{\IEEEauthorblockN{ $^\dag$Vahid Jamali, $^\ddag$Yasser Karimian, \textit{Student Member}, \textit{IEEE}, $^\dag$Johannes Huber, \textit{Fellow}, \textit{IEEE}, \\$^\ddag$Mahmoud Ahmadian, \textit{Member}, \textit{IEEE} 
\\
$^\dag$Friedrich-Alexander-University Erlangen-N\"{u}rnberg (FAU), Erlangen, Germany \\
$^\ddag$K. N. Toosi University of Technology (KNTU), Tehran, Iran}

}

\maketitle
\begin{abstract}
The complexity-performance trade-off is a fundamental aspect of the design of low-density parity-check (LDPC) codes. In this paper, we consider LDPC codes for the binary erasure channel (BEC), use code rate for performance metric, and number of decoding iterations to achieve a certain residual erasure probability for complexity metric. We first propose a quite accurate approximation of the number of iterations for the BEC. Moreover, a simple but efficient utility function corresponding to the number of iterations is developed. Using the aforementioned approximation and the utility function, two optimization problems w.r.t. complexity are formulated to find the code degree distributions. We show that both optimization problems are convex. In particular, the problem with the proposed approximation belongs to the class of \textbf{semi-infinite} problems which are computationally challenging to be solved. However, the problem with the proposed utility function falls into the class of \textbf{semi-definite} programming (SDP) and thus, the global solution can be found efficiently using available SDP solvers. Numerical results reveal the superiority of the proposed code design compared to existing code designs from literature. 
\end{abstract}


\section{Introduction}
\label{Sec: Intro}
Efficient design of low-density parity-check (LDPC) codes under iterative message-passing have been widely investigated in the literature
\cite{Shokrol1,Luby1,Barak,Richardson,Luby2,Hasan}. In \cite{Shokrol1} and \cite{Luby1}, capacity-achieving    ensembles  of LDPC  codes  were  originally  introduced. Several upper bounds on the maximum achievable rate of LDPC codes   over  the  binary  erasure  channel  (BEC)   for a  given  check degree distribution were derived in \cite{Barak}. Moreover, irregular LDPC codes were designed by optimizing the degree structure of the Tanner graphs with code rates extremely close to the Shannon capacity in \cite{Richardson}. These codes are referred to as \textit{performance-optimized codes} since the objective of code design is to obtain degree distributions which maximize the code rate for a given channel as the performance metric. However, for a performance-optimized code, convergence  of  the decoder usually requires a large number of iterations. This leads to high decoding complexity and processing delay which is not appropriate for many practical applications. Hence, the complexity of decoding process has to be considered as a design criterion for the LDPC codes, as well. In contrast to performance-optimized codes, codes which are designed to minimize the decoding complexity for a given code rate are denoted by \textit{complexity-optimized codes}. 

The concept of performance-complexity trade-off under iterative message-passing decoding was introduced in  \cite{KhandekarConf,KhandekarPhD,Sason1,Sason2}. In \cite{KhandekarConf} and \cite{KhandekarPhD}, the authors investigated the decoding and encoding complexity for achieving the channel capacity of the binary erasure channel (BEC). Moreover,  Sason  and  Wiechman \cite{Sason2} showed that the number of decoding iterations which is required to fall below a given residual erasure probability under
iterative  message-passing  decoding,  scales proportional to  the inverse of the gap between code rate and capacity. However, specifically for a code rate significantly below Shannon capacity, there exist different ensembles for the same rate but with different convergence behavior of decoder. Therefore, how to find an ensemble  which leads to the lowest number of decoding iterations is an important question in the code design. 

 Unfortunately, characterizing the number of decoding iterations as a function of code parameters is not straightforward. Hence, several bounds and approximations for the number of  decoding iterations have been proposed in  literature \cite{Smith,XudongISIT,Sason2}. Simple  lower  bounds  on  the number  of  iterations  required  for  successful  message-passing decoding over BEC were proposed in \cite{Sason2}. These bounds are expressed in terms of some basic parameters of the considered code ensemble. In particular, the fraction of degree-2 variable nodes, the target residual erasure probability, and the gap to the channel capacity are considered in \cite{Sason2}. Thereby, the proposed lower bound in \cite{Sason2} suggests that for given code rate, the fraction of degree-2 variable nodes has to be minimized for a low number of decoding iterations. However, all other degree distribution parameters are not included in this lower bound. In \cite{Smith}, an approximation of the number of iterations was proposed and used to formulate an optimization problem for complexity minimization, i.e., a complexity-optimizing problem. The proposed approximation  is a function of all degree distribution parameters. However, the resulting optimization problem is non-convex and is solved iteratively, only. Furthermore, it was proved that under a certain mild condition, the considered optimization problem in each iteration is convex. A similar approach to design a complexity-optimized code was investigated in \cite{XudongISIT}. We note that one of the essential constraints that has to be considered for LDPC code design is to guarantee that the residual erasure probability  decreases after each decoding iteration. In general, this successful decoding constraint for the BEC leads to an optimization problem which belongs to a class of semi-infinite programming, i.e., the problem has infinite number of constraints \cite{Semi-infinite1,Semi-infinite2}. Semi-infinite optimization problems are computationally challenging to be solved.  We will investigate this class of optimization problems in more detail in Section III.

The extrinsic information transfer (EXIT) chart \cite{HuberTr,Kramer} and density evolution \cite{Density} are two powerful tools for tracking the convergence behavior of iterative decoders. For the BEC,  EXIT charts coincide with the density evolution analysis \cite{Sason2,ModernCoding}. For  simple  presentation  of the performance-complexity  tradeoff,  we  analyze  a modified  density  evolution  in this paper. In particular, first a quite accurate approximation of the number of iterations is proposed. 
Based on this approximation, an optimization problem is formulated such that for any given check degree distribution, a variable degree distribution with a finite maximum variable node degree is found such that the number of required decoding iterations to achieve a certain target residual  erasure probability  is minimized while a desired code rate is guaranteed. Although, we prove that the considered optimization problem is convex, it still belongs to the class of \textbf{semi-infinite} programming. Therefore, we first propose a lower bound on the number of decoding iterations. Based on this, a simple, but efficient utility function corresponding to the number of decoding iterations is developed. We show that by applying this utility function, the optimization problem now falls into the class of \textbf{semi-definite} programming (SDP)  where the global solution of the considered optimization problem can be found efficiently using available SDP solvers such as CVX \cite{CVX}. It is worth mentioning that a general framework is developed to prove that the considered problem is SDP. Thus, this framework may be also used to design LDPC codes w.r.t. other design criteria. As an example, we formulate an optimization problem to obtain performance-optimized codes, i.e., for any given check degree distribution, a variable degree distribution with finite maximum variable node degree will be found such that the achievable code rate is maximized. The maximum achievable code rate obtained from the performance-optimizing optimization problem is also used as an upper bound for the desired rate in the considered complexity-optimizing code design.

 To summarize, the contributions of this paper are twofolded: \textit{i}) we propose a quite accurate approximation, a lower bound, and a simple utility function corresponding to the number of decoding iterations and formulate two optimization problems w.r.t. complexity to find the best variable degree distribution for any given check node distribution, and  \textit{ii}) we show that both problems are convex in the optimization variables, and as the main contribution of the paper, we prove that the optimization problem with the proposed utility function has a SDP representability which allows to efficiently solve the problem. Note that the approximation proposed here is derived for the number of iterations between any  two general functions and is quite different from the one proposed in \cite{Smith}. Moreover, the formulated problem using the utility function in this paper is proved to be a semi-definite and convex problem, compared to the semi-infinite problems considered in \cite{Smith} and \cite{XudongISIT}. Numerical results reveal that for a given limited number of iterations, the complexity-optimized codes significantly outperform the performance-optimized codes. Moreover, the codes designed by the proposed utility function require a  number of decoding iterations which is very close to that of the codes designed by the proposed quite accurate approximation.

The rest of the paper is organized as follows: In Section II, some preliminaries of LDPC codes are recapitulated and the proposed approximation of the number of iterations is devised. Section III is devoted to the optimization problems for the design of complexity-optimized codes based on this approximation and the utility function.  The optimization problem for the performance-optimized codes is presented in Section IV. In Section V, numerical results, and comparisons are given and Section VI concludes the paper.

\section{Problem Formulation}

In this section, we first discuss some basic preliminaries of LDPC codes for the BEC. Then, we introduce a modified density evolution and propose an approximation for the number of iterations based on the concept of performance-complexity tradeoff. 

\subsection{Preliminaries}

An ensemble of an irregular LDPC codes is characterized by the edge-degree distributions $\boldsymbol{\lambda}=[\lambda_2,\lambda_3,\dots,\lambda_{d_v}]$ and $\boldsymbol{\rho}=[\rho_2,\rho_3,\dots,\rho_{d_c}]$ with $\sum_{i = 2}^{d_v } {\lambda _i  } = 1$ and $\sum_{i = 2}^{d_c } {\rho _i  } = 1$. In particular, the fraction of edges in the Tanner graph of a LDPC code that are connected to degree-$i$ variable nodes is denoted $\lambda_i$, and the fraction of edges that are connected to degree-$i$ check nodes, is denoted $\rho_i$  (degree distributions from the edge perspective) \cite{Shokrol1}. Moreover, $d_v$ and $d_c$ denote the maximum variable node degree and check  node  degree,  respectively. Furthermore, let
\begin{IEEEeqnarray}{ll}
 \lambda (x) &= \sum\limits_{i = 2}^{d_v } {\lambda _i x^{i - 1} }, \,\, \text{and}\IEEEyesnumber\IEEEyessubnumber\\
 \rho (x) &= \sum\limits_{i = 2}^{d_c } {\rho _i x^{i - 1} } \IEEEyessubnumber
\end{IEEEeqnarray}
be defined as generating functions of the variable and check degree distributions, respectively \cite{Shokrol1}. Using these notation, the rate of the LDPC code  is given by \cite{Sarah}
\begin{IEEEeqnarray}{ll}\label{CodeRate}
    R = 1 - \frac{\displaystyle \int_0^1 {\rho (x)\mathrm{d}x} }{{\displaystyle\int_0^1 {\lambda (x)\mathrm{d}x} }} =
    1- \frac{\displaystyle \sum\limits_{i = 2}^{d_c } {\frac{\rho _i}{i} } }{\displaystyle\sum\limits_{i = 2}^{d_v } {\frac{\lambda _i}{i} } }.
\end{IEEEeqnarray}

We consider a BEC with bit erasure probability $\varepsilon$, i.e., the channel capacity is $C=1-\varepsilon$. Assuming message-passing decoding for an ensemble of LDPC codes with degree distributions $(\boldsymbol{\lambda},\boldsymbol{\rho})$, the average residual erasure probability over all variable nodes at the $l$-th iteration, $P_l$, when the block length tends to infinity, is given by \cite{Sarah}
\begin{equation}\label{ErrorProb}
P_l  = \varepsilon \lambda (1 - \rho (1 - P_{l - 1} )), \quad l=1,2,\dots.
\end{equation}
where $P_0=\varepsilon$. An essential constraint for a successful decoding is  that the residual erasure probability decreases after each decoding iteration, i.e., $P_l < P_{l-1},\,\,\forall l$. In particular,   the  condition to achieve a target residual erasure probability $\eta$ is \cite{Saeedi}
\begin{equation} \label{SuccDec}
  \varepsilon \lambda (1 - \rho (1 - x )) < x, \quad \forall x \in(\eta,\,\,\varepsilon].
\end{equation}
In this paper, we consider the achievable code rate as the performance metric and the number of decoding iterations which yields to a residual average erasure probability below $\eta$ as the complexity metric. Note that the decoding threshold has also been  considered as a performance metric in the literature, i.e., for a given code rate, the best degree distribution is determined such that the successful decoding constraint in (\ref{SuccDec}) holds for  the maximum possible erasure probability $\varepsilon$. However,  the code rate maximization problem for a given erasure probability is in principle equivalent to the threshold maximization problem for a given code rate \cite{Smith}. 

In general, the complexity of the decoding process is comprised of the number of required iterations and the complexity per iteration which is also referred to as graphical complexity. Formally, graphical complexity $\mathcal{G}$ is defined in \cite{Sason2} as the number of edges in the Tanner graph per information bit. In this paper, however, our goal is to find a variable degree distribution for any given check node distribution $\rho(x)$ and code rate $R<C$. Moreover, for given $\rho(x)$, the number of edges in the Tanner graph is obtained as $M=\frac{m}{\int_{0}^{1}\rho(x)\mathrm{d}x}$ where $m$ is the number of check nodes. Furthermore, the code rate $R$ is defined as $R=1-\frac{m}{n}$ where $n$ is the number of variable nodes, i.e., the length of the code. Therefore, for given $\rho(x)$ and $R$,  the graphical complexity is fixed to $\mathcal{G}=\frac{M}{n-m}=\frac{1-R}{R\int_{0}^{1}\rho(x)\mathrm{d}x}$, and hence, not changed by varying $\lambda(x)$. We can conclude that the decoding complexity depends only on the number of decoding iterations. Moreover, the number of iterations is also mainly used to measure the processing delay in decoding \cite{Sason2}. 

\subsection{Number of Decoding Iterations}

\begin{figure}
\centering
\iftoggle{OneColumn}{%
  \pstool[width=0.55\linewidth]
}{%
  \pstool[width=0.9\linewidth]
}
    {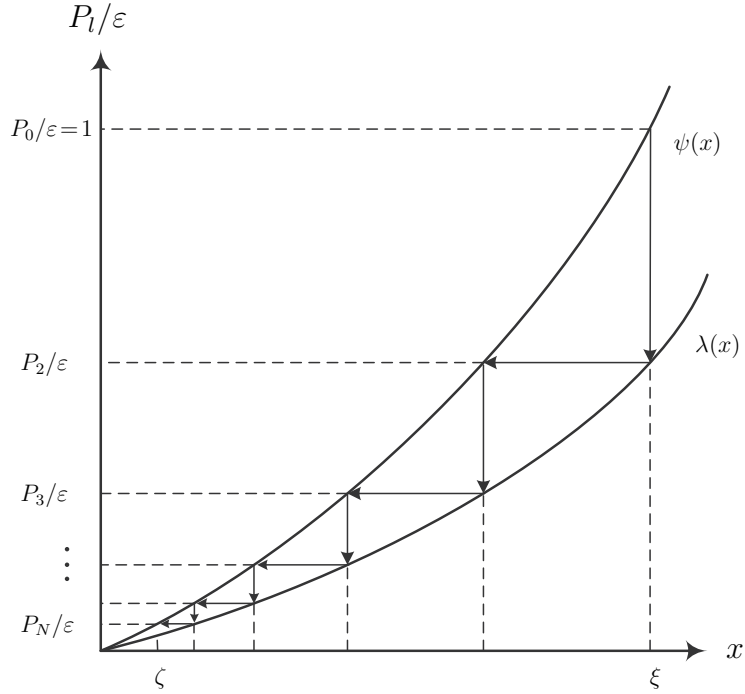}{
    \psfrag{Pl}[t][c][1]{$P_l/\varepsilon$}
\psfrag{P1}[l][c][0.75]{$\, P_0/\varepsilon\Equal 1\,\,$}
\psfrag{P2}[l][c][0.75]{$\,\,\, P_2/\varepsilon$}
\psfrag{P3}[l][c][0.75]{$\,\,\, P_3/\varepsilon$}
\psfrag{Pn}[l][c][0.75]{$\,\,\, P_{N}/\varepsilon$}
\psfrag{S1}[c][c][0.75]{$\,\,\lambda(x)$}
\psfrag{S2}[c][c][0.75]{$\quad\psi(x)$}
\psfrag{X}[c][c][1]{$x$}
\psfrag{Khi}[c][t][0.75]{$\xi$}
\psfrag{Zeta}[c][t][0.75]{$\zeta\,\,\,$}}
    \caption{Modified density evolution for BEC with bit erasure probability $\varepsilon$,  $\psi(x)=\frac{1}{\varepsilon}[1 - \rho^{-1} (1 - x )]$, and $\xi=\psi^{-1}(1)$ and $\zeta=\psi^{-1}(\frac{\eta}{\varepsilon})$.}
    \label{DensityEvol}
\end{figure}

\begin{figure}
\centering
\iftoggle{OneColumn}{%
  \pstool[width=0.55\linewidth]
}{%
  \pstool[width=0.9\linewidth]
}
{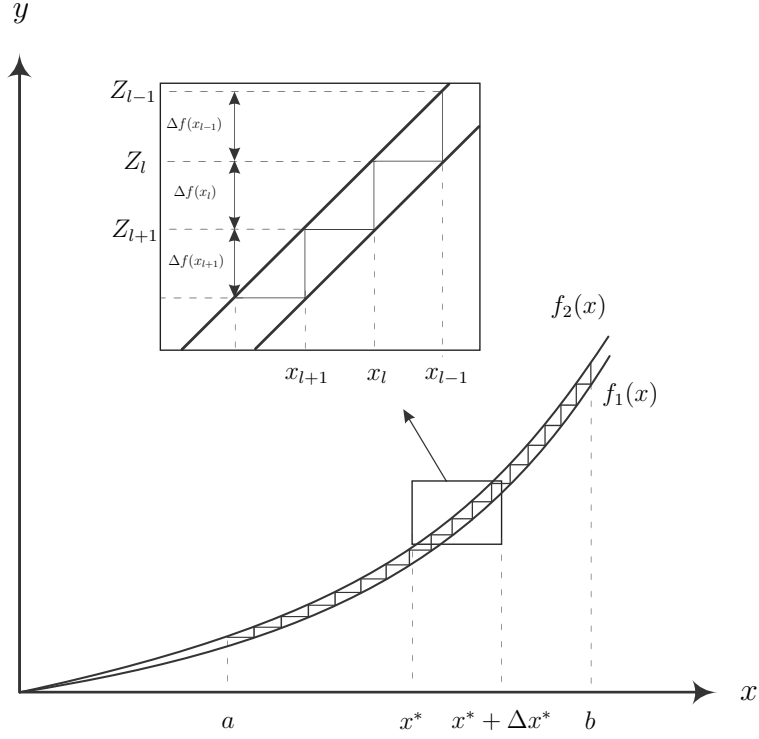}{
    \psfrag{X}[c][c][1]{$x$}
    \psfrag{Y}[c][c][1]{$y$}
\psfrag{X1}[c][c][0.8]{$x_{l-1}$}
\psfrag{X2}[c][c][0.8]{$x_{l}$}
\psfrag{X3}[c][c][0.8]{$x_{l+1}$}
\psfrag{T1}[c][c][0.8]{$Z_{l-1}$}
\psfrag{T2}[c][c][0.8]{$Z_{l}$}
\psfrag{T3}[c][c][0.8]{$Z_{l+1}$}
\psfrag{D1}[c][c][0.45]{$\,\Delta f(x_{l-1})$}
\psfrag{D2}[c][c][0.45]{$\Delta f(x_{l})$}
\psfrag{D3}[c][c][0.45]{$\,\Delta f(x_{l+1})$}
\psfrag{f1}[c][c][0.8]{$f_2(x)$}
\psfrag{f2}[c][c][0.8]{$f_1(x)$}
\psfrag{a}[c][c][0.8]{$a$}
\psfrag{b}[c][c][0.8]{$b$}
\psfrag{xs1}[c][c][0.8]{$x^*$}
\psfrag{xs2}[c][c][0.8]{$x^*+\Delta x^*$}}
    \caption{Approximation of decreasing step from $x_l$ towards $x_{l+1}$ between functions $f_1(x)$ and $f_2(x)$.}
    \label{AppDeriv}
\end{figure}

The expected  performance of a particular ensemble over BEC can be determined by tracking the residual erasure probabilities through the iterative decoding process by density evolution \cite{Sarah,Smith}. In this paper, a modified version of the density evolution is utilized to determine the required number of decoding iterations to achieve a certain residual erasure probability. To this end, we define the following function
\begin{IEEEeqnarray}{ll}\label{SayDef}
  \psi(x)=\frac{1}{\varepsilon}[1 - \rho^{-1} (1 - x )].
\end{IEEEeqnarray}
Note that by defining $\psi(x)$, we have separated the known parameters $\rho(x)$ and $\varepsilon$ from the optimization variables $\lambda(x)$. Then, the residual erasure probability can be tracked by the iterations between the two curves $\lambda(x)$ and $\psi(x)$, cf. eq. (\ref{ErrorProb}) and Fig. \ref{DensityEvol}. In particular, the successful decoding constraint in (\ref{SuccDec}) is equivalent to constraint $\lambda(x)<\psi(x), \,\,x\in(\zeta,\,\,\xi]$ where $\xi=\psi^{-1}(1)=1-\rho(1-\varepsilon)$ and $\zeta=\psi^{-1}(\frac{\eta}{\varepsilon})=1-\rho(1-\eta)$. Furthermore, for a given code rate $R$, we can conclude from (\ref{CodeRate}) that the area bounded by the curves $\lambda(x)$ and $\psi(x)$ is fixed for a given code rate, i.e.
\begin{IEEEeqnarray}{ll}\label{AreaTheo}
  \int_{0}^{1} [\psi(x)-\lambda(x)] \mathrm{d} x = \left(\frac{1}{\varepsilon}-\frac{1}{1-R}\right)\int_{0}^{1}\rho(x)\mathrm{d}x.
\end{IEEEeqnarray}
The above property is well known as area Theorem \cite{Kramer}.  Therefore, the problem of code design can be interpreted as a curve shaping problem for $\lambda(x)$ subjected to the constraints that $\lambda(x)$ is below $\psi(x)$ and the area bounded by the two curves is fixed. Asymptotically, as the target code rate approaches the capacity, $C-R\to 0$, the area bounded by the curves $\lambda(x)$ and $\psi(x)$ vanishes, and the number of decoding iterations grows to infinity. However, for the code rate below the capacity, one can find different $\lambda(x)$ that have code rate $R$ while acquiring different convergence behaviors. Therefore, the goal of code design is to shape the curves $\lambda(x)$ for the best complexity-performance trade-off. In the following, we utilize the distance concept of performance-complexity trade-off  illustrated in Fig. \ref{DensityEvol} to propose a quite accurate approximation for the number of iterations. To this end, we state the following definition, see also Fig. \ref{AppDeriv}.

\begin{defin}
Let functions $f_1(x)$ and $f_2(x)$ have positive first-order derivatives and satisfy condition $f_1(x)<f_2(x),\,\,x\in[a,\,\,b]$. Then, $N(f_1(x),f_2(x),a,b)$ is defined as 
\begin{IEEEeqnarray}{ll}\label{NumberDef}
  N(f_1(x),f_2(x),a,b) = \min\underset{l}{\arg}\{Z_l<a\}
\end{IEEEeqnarray}
where
\begin{IEEEeqnarray}{ll}\label{FuncTl}
  Z_l=f_1(f_2^{-1}(Z_{l-1})),\quad\text{and}\quad Z_0=b.
\end{IEEEeqnarray}
\end{defin}

From the above definition, the number of iterations to achieve a certain target residual erasure probability $\eta$ is given by $N(\lambda(x),\psi(x),\eta, \varepsilon )$. We observe that due to iterative structure of the decoding process, the number of iterations as a function of code parameters is a non-differentiable function which is difficult to deal with in a code design in general and does not offer much insight. In the following, we propose a continuous approximation of the number of iterations between two general functions $f_1(x)$ and $f_2(x)$.

We first define a function for the distance between $f_2(x)$ and $f_1(x)$, i.e.,  $\Delta f(x) = f_2(x) -f_1(x)$, which plays a main role in
characterizing the number of iterations, i.e., $\Delta f(x)$ is the decreasing step at $x$. Then, we can rewrite $Z_l$ in (\ref{FuncTl}) as follows
\begin{equation}
 Z_l = Z_0 - \sum\limits_{i=0}^{l-1} \Delta f(x_i),
\end{equation}
where $Z_l = f_2(x_l)$. Assume 
 $x\in[x^*,\,\,x^*+ \Delta x^*]\subset[a,\,\,b]$ where $\Delta x^*$ is sufficiently small such 
that $\Delta f(x)$ is assumed to be constant within this interval, see Fig. \ref{AppDeriv}.
Thus, each decreasing step in this interval toward axis $y$, i.e., $\Delta f(x)$, or toward axis $x$, i.e., $\Delta x=x_l - x_{l+1}$, is fixed with the relation $\Delta f(x)=f_2'(x)\Delta x$. Hence, the number of iterations in this interval is given by
\begin{equation}\label{Line}
 \Delta N(f_1(x),f_2(x),x^*,x^* \Add \Delta x^*) \Equal \left\lceil \frac{\Delta x^*}{\Delta x} \right\rceil \Equal \left\lceil \frac{f_2'(x)\Delta x^*}{f_2(x)\Minus f_1(x)} \right\rceil.
\end{equation}
where $\lceil x \rceil = \min \{m\in\mathbb{Z}|m\geq x\}$ is the ceiling function where $\mathbb{Z}$ is the set of integer numbers. The main idea of extending the above expression to all interval is to momentarily ignore the fact that the number of iterations has to be an
integer,  and  to  compute  the  incremental  increase  in the number of iterations as
a  function of the  incremental change in $x$. Notice that a similar  method was also used in \cite{Smith} but a completely different approximation had been obtained. By this, the incremental change in the number of iterations as a function of incremental change in $x$ can be written as $\Delta N = \frac{f_2'(x)\Delta x}{f_2(x)-f_1(x)}$. 
To calculate the total number of iterations, we use the integration over $[a,\,\,b]$ which leads to
\iftoggle{OneColumn}{%
\begin{IEEEeqnarray}{lcl}\label{Approx}
  N(f_1(x),f_2(x),a,b)  \approx& \int_a^b \frac{f_2'(x)}{f_2(x)-f_1(x)} \mathrm{d}x 
  \overset{(\mathrm{here})}{=}& \int_{\zeta}^{\xi} \frac{\psi'(x)}{\psi(x)-\lambda(x)} \mathrm{d}x
\end{IEEEeqnarray}
}{%
\begin{IEEEeqnarray}{lcl}\label{Approx}
  N(f_1(x),f_2(x),a,b)  &\approx& \int_a^b \frac{f_2'(x)}{f_2(x)-f_1(x)} \mathrm{d}x \nonumber \\
  &\overset{(\mathrm{code\, design})}{=}& \int_{\zeta}^{\xi} \frac{\psi'(x)}{\psi(x)-\lambda(x)} \mathrm{d}x
\end{IEEEeqnarray}
}
where equality $(\mathrm{code\, design})$ is obtained as an approximation for the number of decoding iterations, i.e., by substituting $f_1(x)=\psi(x)$, $f_2(x)=\lambda(x)$, $a=\zeta$, and $b=\xi$. Since we use the continuous assumption in the incremental change of the number of iterations, the expression in (\ref{Approx}) is  an approximation of the number of iterations.

In Section IV, it is shown that the approximation (\ref{Approx}) of number of iterations is quite accurate for several numerical examples. Moreover, this approximation  has nice properties compared to the number of iterations given in (\ref{NumberDef}): \textit{i)} differentiability, and \textit{ii)} convexity w.r.t. the optimization variables $\lambda_i$. The convexity of (\ref{Approx})  will be investigated in detail in the following section.

\section{Complexity-Optimized LDPC Codes}
In this section, we discuss the approximation (\ref{Approx})  for the number of decoding iterations for the complexity-optimizing code design.  Moreover, to facilitate obtaining a complexity-optimized code, another optimization problem is also formulated based on a utility function corresponding to the number of decoding iterations.  

\subsection{Code Design with Approximation (\ref{Approx})}
In this subsection, we formulate an optimization problem to find a fast convergent LDPC code based on the approximation in (\ref{Approx}) for BEC with bit erasure probability $\varepsilon$, target residual erasure probability $\eta$, and desired code rate $R_d$. In particular, the following optimization problem is considered to obtain the complexity-optimized LDPC codes
\begin{IEEEeqnarray}{ll}\label{ProbApprox}  
  {\underset{\boldsymbol{\lambda}}{\mathrm{minimize}}} \quad & \int_{\zeta}^{\xi} \frac{\psi'(x)}{\psi(x)-\lambda(x)} \mathrm{d}x \nonumber \\
  \mathrm{subject \,\, to} & \mathrm{C1:}  \lambda(x) < \psi(x), \quad \forall x \in [\zeta,\,\, \xi] \nonumber \\
   &\mathrm{C2:}  R \geq R_d \nonumber \\
   &\mathrm{C3:}  \sum\limits_{i = 2}^{d_v } {\lambda _i } = 1  \nonumber \\
  &\mathrm{C4:}   \lambda _i \geq 0, \quad i=2,\dots,d_v,
\end{IEEEeqnarray}
where $\psi(x)$ is defined in (\ref{SayDef}). The cost function is indeed the approximation (\ref{Approx}), constraint $\mathrm{C1}$ is the decoding constraint in (\ref{SuccDec}), constraint $\mathrm{C2}$ specifies the code rate requirement, and constraints $\mathrm{C3}$ and $\mathrm{C4}$ impose the  restrictions of the definition of degree distribution $\boldsymbol{\lambda}$. The following lemma states that the problem in (\ref{ProbApprox}) is convex.

\begin{lem}
The optimization problem (\ref{ProbApprox}) is convex w.r.t optimization variables $\boldsymbol{\lambda}$ and in domain $x\in (\zeta,\,\,\xi] $.
\end{lem}

\begin{IEEEproof}
We first note that constraints $\mathrm{C1}$, $\mathrm{C3}$, and $\mathrm{C4}$ in the optimization problem (\ref{ProbApprox}) are affine in $\boldsymbol{\lambda}$. Moreover, constraint $\mathrm{C2}$ can be rewritten as $ \sum\limits_{i = 2}^{d_v } {\frac{\lambda _i}{i} } = \frac{1 }{1-R_d } \sum_{i = 2}^{d_c } {\frac{\rho _i}{i} }$ which is an affine form in $\boldsymbol{\lambda}$. Therefore, it suffices to show the convexity of the cost function $\int_{\zeta}^{\xi} \frac{\psi'(x)}{\psi(x)-\lambda(x)} \mathrm{d}x$. Moreover, the convexity of the integrand is sufficient for the convexity of the integral since integration preserves convexity \cite{Boyd}.  To show the convexity of the integrand, i.e., $f(\boldsymbol{\lambda})=\frac{\psi'(x)}{\psi(x)-\lambda(x)}$, we show that it has a positive semi-definite Hessian matrix, i.e., $\nabla^2 f(\boldsymbol{\lambda}) \succeq 0$, which is a sufficient condition of convexity \cite{Boyd}. In particular, the Hessian matrix of $f(\boldsymbol{\lambda})$ is given by
\begin{IEEEeqnarray}{ll}\label{Hessian}
\nabla^2 f(\boldsymbol{\lambda}) = \frac{2\psi'(x)}{\left(\psi(x)-\lambda(x)\right)^3} 
\begin{bmatrix}  x^2 &   x^3 &   \cdots  &  x^{d_v}  \\
 x^3 &   x^4 &   \cdots  & x^{d_v+1} \\
 \vdots & \vdots & \ddots & \vdots \\
 x^{d_v} & x^{d_v+1} & \cdots & x^{2d_v-2} \end{bmatrix} \quad
\end{IEEEeqnarray}
A matrix is positive semi-definite if all its eigenvalues are non-negative. Moreover, the trace of a matrix is equal to the sum of its eigenvalues, i.e., $\mathrm{tr}\{\nabla^2 f(\boldsymbol{\lambda})\}=\sum_{i=1}^{d_v-1} \upsilon_i$ holds where $\upsilon_i,\,\,i=1,\dots,d_v-1$ are the eigenvalues of $\nabla^2 f(\boldsymbol{\lambda})$. Herein, the Hessian matrix in (\ref{Hessian}) has rank one since all the columns of the matrix is linearly dependent with the first column. This implies that all the eigenvalues are zero except one.  The non-zero eigenvalue is given by
\begin{IEEEeqnarray}{ll}
\upsilon = \frac{2\psi'(x)}{\left(\psi(x)-\lambda(x)\right)^3} 
\sum\limits_{i=1}^{d_v-1} x^{2i}>0
\end{IEEEeqnarray}
where we use  $\psi'(x)\geq 0$ and $\lambda(x)<\psi(x),\,\,x\in (0\,\,\xi] $. Thus, the Hessian matrix of the integrand is positive semi-definite and the integrand and consequently the cost function in (\ref{ProbApprox}) are convex. This completes the proof.
\end{IEEEproof}

Although the optimization problem in (\ref{ProbApprox}) is convex, it still belongs to the class of semi-infinite programming \cite{Semi-infinite1,Semi-infinite2}. In particular, optimization problems with finite number of variables and infinite number of constraints or alternatively, infinite number of variables and finite number of constraints are referred to as semi-infinite programming. Herein, the optimization problem in (\ref{ProbApprox}) has finite number of variables, i.e., $\lambda_i,\,\,i=2,\dots,d_v$, but infinite number of constraints since $\lambda(x) < \psi(x)$ must hold for all $x \in (\zeta,\,\, \xi]$. One way to solve the optimization problem in (\ref{ProbApprox})  is to approximate the continuous interval $x \in (\zeta,\,\, \xi]$ by a discrete set $\mathcal{X}=\{x_0,x_1,\dots,x_n\}$ \cite{Semi-infinite2}. We note that a discrete set is also required to numerically calculate the integral expression for the number of iterations since, in general, no closed form solution is available. Thereby, we have finite number of constraints $\lambda(x_i) < \psi(x_i),\,\,i=1,\dots,n$ and the cost function can be expressed as $\sum\limits_{i=2}^{n-1} \frac{\psi'(x_i)\Delta x_i}{\psi(x_i)-\lambda(x_i)}$ where  $\Delta x_i = \frac{x_{i+1}-x_{i-1}}{2}$. For a uniform discretization, we obtain  $\Delta x_i=\frac{\xi-\zeta}{n}\,\,\forall i$, $x_i=x_{i-1}+\Delta x$, $x_0=\zeta$, and $x_n=\xi$.

The solution obtained  via this discritization method is asymptotically optimal as $n\to\infty$. However, it is in general computationally challenging. The problems associated with the successful decoding constraint for BEC given in (\ref{SuccDec}) encounters the same computational complexity, e.g., such as the problems considered in \cite{Smith,XudongISIT}.  

\subsection{Code Design by Means of a Utility Function}

In this subsection, our goal is to formulate and solve an optimization problem based on a utility function corresponding to the number of decoding iterations. In particular, a lower bound on the number of iterations is first proposed. Based on this lower bound, we develop this utility function and show that the resulting optimization problem  can be solved more efficiently than the optimization problem (\ref{ProbApprox}).  Optimizing a utility function instead of the original problem has been frequently applied in practical and engineering designs when the original problem is not manageable or computationally challenging (NB: using pair-wise error probability (PEP) instead of the exact error probability or using minimum min square error (MMSE) and zero-forcing (ZF) detectors instead of maximum-likelihood detector are the well-known examples).

\begin{lem}\label{Lower}
For two functions $f_1(x)$ and $f_2(x)$ with positive first-order derivatives and satisfying $f_1(x)<f_2(x)$ in interval $[a,\,\,b]$, cf. Fig. \ref{AppDeriv}, where the area bound by the two functions is fixed, i.e., $\int_{a}^{b} f_2(x)-f_1(x) \mathrm{d}x = c$, the approximation of the number of iterations is lower bounded by
\begin{IEEEeqnarray}{ll}\label{LowerBound}
  \int_{a}^{b} \frac{f_2'(x)}{f_2(x)-f_1(x)} \mathrm{d}x \geq \displaystyle \frac{b-a}{c}\left( f_2(b)-f_2(a)\right),
\end{IEEEeqnarray}
where the inequality holds with equality if
\begin{IEEEeqnarray}{ll}\label{LowerCondition}
  f_1(x) = f_2(x) - \frac{c}{f_2(b)-f_2(a)} f_2'(x), \qquad x\in[a,\,\,b].\,\,\,
\end{IEEEeqnarray}
\end{lem}

\begin{IEEEproof}
The Jensen's inequality is used to obtain the lower bound. In particular, for a convex function $\varphi(\cdot)$ and a non-negative function $f(x)$ over $[a,\,\,b]$, Jensen's inequality indicates
\begin{IEEEeqnarray}{ll}\label{JensenOrg}
  \varphi\left(\int_{a}^{b} f(x) \mathrm{d}x \right) \leq \displaystyle \frac{1}{b-a} \int_{a}^{b} \varphi\left( (b-a)f(x)\right) \mathrm{d}x,
\end{IEEEeqnarray}
where the inequality holds with equality if and only if $f(x)=d,\,\,x\in[a,\,\,b]$ where $d$ is a constant. In order to use the Jensen's inequality, we assume $\varphi(\gamma)=\frac{1}{\gamma},\,\,\gamma>0$ and $f(x)=\frac{f_2(x)-f_1(x)}{f_2'(x)}$. Therefore, the following lower bound for the approximation (\ref{Approx}) is obtained
\begin{IEEEeqnarray}{ll}\label{Jensen}
  \int_{a}^{b} \frac{f_2'(x)}{f_2(x)-f_1(x)} \mathrm{d}x \geq \displaystyle \frac{(b-a)^2}{\int_{a}^{b} \frac{f_2(x)-f_1(x)}{f_2'(x)} \mathrm{d}x},
\end{IEEEeqnarray}
where the lower bound in  (\ref{Jensen}) is achieved  with equality if and only if
\begin{IEEEeqnarray}{ll}\label{Struct}
   \frac{f_2(x)-f_1(x)}{f_2'(x)}\Equal d \,\,\text{or}\,\, f_1(x) = f_2(x) \Minus d f_2'(x), \,\, x\in [a,\,\,b].\quad\,\,
\end{IEEEeqnarray}
In order to obtain constant $d$, we integrate both side of $f_2(x) - f_1(x) = d f_2'(x)$ over the interval $[a,\,\,b]$ which leads to
\begin{IEEEeqnarray}{ll}\label{dOpt}
   d = \frac{c}{f_2(b)-f_2(a)} \cdot
\end{IEEEeqnarray}
Substituting $d$ in (\ref{dOpt}) into (\ref{Struct}) and (\ref{Jensen}) gives the lower bound (\ref{LowerBound}) and the condition (\ref{LowerCondition}) to achieve this lower bound with equality stated in Lemma \ref{Lower}. Notice that the constant $d$ corresponds to a decent in Figs. \ref{DensityEvol} and \ref{AppDeriv} with equal length steps on abscissa $x$. We conclude from (\ref{LowerBound}) and (\ref{LowerCondition}) that the number of iterations is minimized if $f_1(x)$ is chosen in a way that, for a given area $c$, all the steps have equal length. This completes the proof. 
\end{IEEEproof}

For a given $f_2(x)$, the choice of $f_1(x)$ introduced in (\ref{LowerCondition}) to achieve the lower bound with equality provides an insightful design tool. Specifically, we can conclude that the best choice of $f_1(x)$  for any given $f_2(x)$ is the one that has the maximum distance with $f_2(x)$ weighted by $f'_2(x)$. For the number of decoding iterations, we have to set $f_1(x)=\lambda(x),f_2(x)=\psi(x),a=\zeta$, $b=\xi$. Then, we propose the smallest step size as a utility function, i.e.
\begin{IEEEeqnarray}{lll}\label{Utility}
  U(\boldsymbol{\lambda})={\underset{x \in [\tilde{\zeta},\,\, \xi] }{\min}}\left\{\frac{\psi(x)-\lambda(x)}{\psi'(x)}\right\}.
\end{IEEEeqnarray}
where $\tilde{\zeta}$ is a design parameter with a value close to $\zeta$. $U(\boldsymbol{\lambda})$ denotes the smallest step size of decent on abscissa $x$ for all $x\in [\tilde{\zeta},\,\, \xi]$, i.e., intuitively the bottleneck within the iterative process. The lower bound (\ref{LowerBound}) indicates that there should not exists any such bottleneck. Thus, a maximization of $U(\boldsymbol{\lambda})$ is obviously reasonable for lowering the number of iterations.

\begin{remk}
Note that for the constraints in Lemma \ref{Lower} on $f_1(x)$, i.e., $f_1(x)<f_2(x)$ and $\int_{a}^{b} f_2(x)-f_1(x) \mathrm{d}x = c$, the optimal choice of $f_1(x)$ to minimize the number of iterations $N(f_1(x),f_2(x),a,b)$ is already given in (\ref{LowerCondition}) and there is no need for optimization. However, for the code design, we have an extra  constraint on $\lambda(x)$  which is the structure of $\lambda(x)$, i.e., $\sum_{i = 2}^{d_v } {\lambda _i } = 1$ and $\lambda _i \geq 0, \,\, i=2,\dots,d_v$. Therefore, (\ref{LowerCondition}) is not directly applicable for the code design. However, as we will see in Section V, the insight that (\ref{LowerCondition}) offers, i.e., maximizing the smallest step size $U(\boldsymbol{\lambda})$, is very efficient and leads to codes with quite low number of decoding iterations.
\end{remk}

\begin{remk}
The reason that we do not choose $\tilde{\zeta}=\zeta$ is that the expression in (\ref{Utility}) is a utility function corresponding to the number of iterations and is neither the exact nor an approximation of the number of iterations. Hence, the choice of $\tilde{\zeta}=\zeta$ does not necessarily lead to the minimum number of decoding iterations at the target residual erasure probability $\eta$, note that $\zeta=\psi^{-1}(\frac{\eta}{\varepsilon})$. For the code design, we can choose $\tilde{\zeta}$  for the lowest number of iterations at the target residual erasure probability.
\end{remk}

Now, we are ready to find the complexity-optimized codes by the maximizing the minimum step size, i.e., the utility function in (\ref{Utility}), for a given desired code rate $R_d$, as follows
\begin{IEEEeqnarray}{lll}\label{Op1}
  {\underset{\boldsymbol{\lambda}}{\mathrm{maximize}}} \quad & U(\boldsymbol{\lambda})\nonumber \\
  \mathrm{subject \,\, to} & \mathrm{C1:}  \lambda(x)< \psi(x), \quad \forall x \in [\tilde{\zeta},\,\, \xi], \nonumber \\
   &\mathrm{C2:}  R \geq R_d, \nonumber \\
   &\mathrm{C3:}  \sum\limits_{i = 2}^{d_v } {\lambda _i } = 1, \nonumber \\
  &\mathrm{C4:}   \lambda _i \geq 0, \quad i=2,\dots,d_v.
\end{IEEEeqnarray}
In order to show that the optimization problem (\ref{Op1}) is a semi-definite programming, we introduce an auxiliary variable $t$ and maximize $t$ where $U(\boldsymbol{\lambda})\geq t\geq 0$ holds as a constraint. Moreover, considering $\psi'(x)>0,\,\,x\in[\tilde{\zeta},\,\,\xi]$, we can rewrite $U(\boldsymbol{\lambda})\geq t$ as $ \psi(x)-\lambda(x) \geq t \psi'(x),\,\,x\in[\tilde{\zeta},\,\,\xi]$ which leads to the following equivalent optimization problem
\begin{IEEEeqnarray}{lll}\label{Op2}  {\underset{\boldsymbol{\lambda},t}{\mathrm{maximize}}} \,\,\, & \quad t  \nonumber \\
  \mathrm{subject\,\, to} & \mathrm{C1}:   \psi(x)-\lambda(x) \geq t \psi'(x),\quad \forall x \in (\tilde{\zeta},\,\, \xi] \nonumber \\
  & \mathrm{C2}:  \sum\limits_{i = 2}^{d_v } {\frac{\lambda _i}{ i} } \geq \frac{1}{1-R_d}\sum\limits_{i = 2}^{d_c } {\frac{\rho _i}{ i} }  \nonumber \\
  & \mathrm{C3}: \sum\limits_{i = 2}^{d_v } {\lambda _i } = 1 \nonumber \\
  & \mathrm{C4}:  \lambda _i,t \geq 0, \quad i=2,\dots,d_v
\end{IEEEeqnarray}

The above optimization problem is still a semi-infinite programming  since it contains infinite number of constraints with respect to $\forall x \in [\tilde{\zeta},\,\, \xi]$. In the following, we  state a lemma which is useful to transform some category of semi-infinite problems into equivalent SDP problems.

\begin{lem}\label{Nemirov}
 Let $\pi(x) = \pi_0 + \pi_1 x + \dots + \pi_{2k} x^{2k}$ be a polynomial of degree $2k$. There exists a matrix $\boldsymbol{\Lambda}$ where a constraint $\pi(x)\geq 0,\,\,\forall x \in \mathbb{R}$ has the following SDP representability w.r.t. variable $\pi_i,\,\,i=0,\dots,2k$
\begin{IEEEeqnarray}{ll}
   \exists \boldsymbol{\Lambda},
\quad  \pi(x) \geq 0, \,\, \forall x \in \mathbb{R}, \,\, \Longleftrightarrow \,\,
   \begin{cases}
   \boldsymbol{\Lambda} \succeq 0, \\
   \sum\limits_{m + n = i+2} { \Lambda_{mn} } = \pi_i,
   \end{cases}
\end{IEEEeqnarray}
where $\Lambda_{mn}$ is the element in $m$-th row and $n$-th column of the matrix $\boldsymbol{\Lambda}$.
\end{lem}
\begin{IEEEproof}
Please refer to \cite[Chapter 4]{Nemirovski}.
\end{IEEEproof}

For the clarity of Lemma \ref{Nemirov}, we consider as an example the quadratic polynomial $\pi(x)=ax^2+bx+c$. Then, from Lemma \ref{Nemirov}, we obtain an equivalent SDP representation of $\pi(x)\geq 0,\,\,x \in \mathbb{R}$ as $\begin{bmatrix}  \Lambda_{11} &   \Lambda_{12}  \\ \Lambda_{21} &  \Lambda_{22} \end{bmatrix} \succeq 0$ where $\Lambda_{11}=\pi_0=c$, $\Lambda_{12}+\Lambda_{21}=\pi_1=b$, and $\Lambda_{22}=\pi_2=a$. The aforementioned SDP representation is equivalent to the well-known conditions $b^2-4ac\leq 0$ and $a>0$ for the non-negativity of a quadratic polynomial. 

Note that Lemma \ref{Nemirov} is developed for $x \in \mathbb{R}$ and variables $\pi_i,\,\,i=0,\dots,2k$. Therefore, to apply Lemma \ref{Nemirov} to the first constraint in (\ref{Op2}), we have to consider: \textit{i}) the interval $[\tilde{\zeta},\,\, \xi]$ has to be mapped to
$\mathbb{R}$, and \textit{ii}) the coefficients of the polynomial functions in constraint $\mathrm{C1}$ in  (\ref{Op2}) have to be calculated and shown to be in an affine form in the optimization
parameters ($\boldsymbol{\lambda},t$). In the following theorem, we present the SDP representation of constraint $\mathrm{C1}$ in (\ref{Op2}). To this end, the function $\psi(x)$ is expanded into a Taylor series $\psi(x) = {\underset{M\rightarrow \infty}{\textrm{lim}}} \sum_{i = 2}^M {T_i x^{i - 1} }$ around $x=0$. 

\begin{theo}\label{SDP}
The constraint $\pi(x)=\psi(x)-\lambda(x)-t\psi'(x)\geq 0,\,\,\forall x \in [\tilde{\zeta},\,\, \xi]$, i.e., constraint $\mathrm{C1}$ in (\ref{Op2}), has the following equivalent SDP representation
\begin{IEEEeqnarray}{CC}
   \begin{cases}
   \boldsymbol{\Lambda} \succ 0, \\
   \sum\limits_{m + n = k+2} { \Lambda_{mn} } = \pi_k,\,\,\,\,\,\, 0\leq k \leq D
   \end{cases}
\end{IEEEeqnarray}
where $\pi_0=0$ and
\iftoggle{OneColumn}{%
\begin{IEEEeqnarray}{ll}
\pi_k = \sum\limits_{i = 1}^{D}{ f_i {\tilde{\zeta}}^i \left( \sum\limits_{j = 0}^{k}{a_{k - j}b_j}\right)} 
 - \sum\limits_{i = 1}^{D}{ iT_i {\tilde{\zeta}}^{i-1} \left( \sum\limits_{j = 0}^{k}{c_{k - j}d_j}\right)} t,\quad k=1,\dots, D
\end{IEEEeqnarray}
}{%
\begin{IEEEeqnarray}{ll}
\pi_k &= \sum\limits_{i = 1}^{D}{ f_i {\tilde{\zeta}}^i \left( \sum\limits_{j = 0}^{k}{a_{k - j}b_j}\right)} \nonumber \\
&\quad - \sum\limits_{i = 1}^{D}{ iT_i {\tilde{\zeta}}^{i-1} \left( \sum\limits_{j = 0}^{k}{c_{k - j}d_j}\right)} t,\quad k=1,\dots, D
\end{IEEEeqnarray}
}
where $D=M-1$ and coefficients $f_i, a_j, b_j,c_j$ and $d_j$ are given by
\begin{IEEEeqnarray}{CCll}\label{Coefficients}
   f_i &=&
   {\begin{cases}
   T_{i+1} -  \lambda_{i+1},  & i = 1,\dots, d_v - 1, \\
   T_{i+1}, & \mathrm{ otherwise }
   \end{cases}} \IEEEyesnumber \IEEEyessubnumber \\
 a_j &=&
   {\begin{cases}
   {i \choose j} \frac{\xi^j}{\tilde{\zeta}^j},  &  0\leq j\leq i \\
   0, & \mathrm{otherwise }
   \end{cases}} \IEEEyessubnumber \\
      b_j &=&
   {\begin{cases}
   {D-i \choose j} ,  &  0\leq j\leq D-i \\
   0, & \mathrm{otherwise }
   \end{cases}} \IEEEyessubnumber \\
    c_j &=&
   {\begin{cases}
   {i-1 \choose j} \frac{\xi^j}{\tilde{\zeta}^j},  &  0\leq j\leq i-1 \\
   0, & \mathrm{otherwise }
   \end{cases}} \IEEEyessubnumber \\
      d_j &=& 
   {\begin{cases}
   {D-i+1 \choose j} ,  &  0\leq j\leq D-i+1 \\
   0, & \mathrm{otherwise }
   \end{cases}}. \IEEEyessubnumber 
\end{IEEEeqnarray}
\end{theo}
\begin{IEEEproof}
Please refer to Appendix \ref{AppSDP}.
\end{IEEEproof}

We note that the matrix elements, i.e., $\Lambda_{mn}$, are in an affine form of the optimization variables, since all $\pi_i$ are affine in $t$ and $f_i$ and finally $f_i$ is affine in $\lambda_{i+1}$. Therefore, all the constraints of the optimization problem in (\ref{Op2}) have shown to be affine or matrix semi-definite constraints. Thus, the optimization problem in (\ref{Op2}) is SDP and can be efficiently solved using  available SDP solvers \cite{CVX}.

\begin{remk}
Note that, for practical code design, a relatively large $M$ is enough for a quite accurate approximation $\psi(x) \approx  \sum\nolimits_{i = 2}^M {T_i x^{i - 1} }$. Moreover, Taylor series coefficients $T_i$ can be represented in close form for check regular ensembles, i.e., $\rho(x)=x^{d_c-1}$, as
\begin{equation}
   T_i =  {\omega \choose  i - 1} \left( { - 1} \right)^i ,\,\,\omega = 1/\left( {d_c  - 1} \right)
\end{equation}
where ${\omega \choose  i }$ is the fractional binomial expansion and defined for real valued $\omega$ and a positive integer valued $i$ as \cite{Shokrol2,Saeedi}
\begin{equation}
   {\omega \choose  i } = \frac{\omega(\omega-1)\dots (\omega-i+1)}{i!}
\end{equation}

\end{remk}

\section{Performance-Optimized LDPC Code}

A general framework has been developed for the statement and the proof of Theorem \ref{SDP} such that it can be possibly used to formulate and solve optimization problems with similar structures. In particular, the optimization problems that have constraint like $f(x,\boldsymbol{\lambda})\geq 0$ where $\boldsymbol{\lambda}$ contains  the optimization variables and the constraint must hold for all $x$ within an interval $[a,\,\,b]$ might have equivalent SDP representation as shown by the framework of Section III and Appendix \ref{AppSDP}. Specifically, one should first map the interval $[a,\,\,b]$ to $(-\infty,\,\,+\infty)$ and then show that the coefficients of the resulting polynomial are in affine form of the optimization variables. As a relevant example, we formulate an optimization problem for code rate maximization to obtain the performance-optimized code in the following and show how the global optimal solution can be obtained via the optimization framework developed in this paper.

For a performance-optimized code, our goal is to maximize the code rate for BEC with given bit erasure probability $\varepsilon$ such that the successful decoding is guaranteed. This optimization problem is formulated as follows
\begin{IEEEeqnarray}{lll}\label{Rate1}
  {\underset{\boldsymbol{\lambda}}{\mathrm{maximize}}} \quad & R \nonumber \\
  \mathrm{subject \,\, to} & \mathrm{C1:}  \lambda(x)<\psi(x), \quad \forall x \in (0,\,\, \xi], \nonumber \\
   &\mathrm{C2:}  \sum\limits_{i = 2}^{d_v } {\lambda _i } = 1, \nonumber \\
  &\mathrm{C3:}   \lambda _i \geq 0, \quad i=2,\dots,d_v.
\end{IEEEeqnarray}
It can easily be observed that constraints $\mathrm{C2}$ and $\mathrm{C3}$  are affine in the optimization variables $\boldsymbol{\lambda}$ and constraint $\mathrm{C1}$ has a SDP representation in the optimization variables $\boldsymbol{\lambda}$ using the aforementioned Taylor series expansion of $\psi(x)$. Moreover, from (\ref{CodeRate}), we can conclude that maximizing the code rate for a given $\rho(x)$ is equivalent to maximizing $\sum_{i = 2}^{d_v }\frac{\lambda _i }{i}$. Therefore, with a similar approach as in problem (\ref{Op1}), we can write the following SDP representation of (\ref{Rate1})
\begin{IEEEeqnarray}{lll}\label{Rate2}  
{\underset{\boldsymbol{\lambda}}{\mathrm{maximize}}} \,\,\, & \quad \sum\limits_{i = 2}^{d_v } \frac{\lambda _i }{i}  \nonumber \\ \vspace{-0.2cm}
  \mathrm{subject\,\, to} & \mathrm{C1}:      \boldsymbol{\Lambda} \succ 0,  \sum_{m + n = k+ 2} {\hspace{-0.3cm} \Lambda_{mn} } = \pi_k,\,\, k=0,\dots,D \nonumber \\ 
  & \mathrm{C2}: \sum\limits_{i = 2}^{d_v } {\lambda _i } = 1 \nonumber \\
  & \mathrm{C3}:  \lambda _i \geq 0, \quad i=2,\dots,d_v
\end{IEEEeqnarray}
where $ \Lambda_{mn}, \pi_k$, and $D$ are the same as the ones given in Theorem \ref{SDP} setting $t=0$ and $\tilde{\zeta}=0$. 

\begin{remk}
The performance-optimized codes resulting from (\ref{Rate2}) are not constrained w.r.t. the required number of decoding iterations. Thus, the the achievable code rate of the performance-optimized code, $R_{\mathrm{max}}$, can be used as an upper bound for the rate constraint in the complexity-optimizing problems. In other words, the desired rate $R_d$ for the complexity-optimized code has to be chosen such that $R_d \leq R_{\mathrm{max}}$ holds, otherwise, the optimization problems in (\ref{ProbApprox}) and (\ref{Op1}) become infeasible. 
\end{remk}

\begin{remk}
We note that the maximum achievable code rate obtained from (\ref{Rate2}) usually is below the channel capacity, $R_{\mathrm{max}}\leq C=1-\varepsilon$, since a finite maximum variable degree $d_v$ is assumed. In particular, one can conclude from (\ref{AreaTheo}) that as $R\to C$, the area between curves $\lambda(x)$ and $\psi(x)$ vanishes which leads to $\lambda(x)\to \psi(x),\,\,  x \in (0\,\, \xi]$. In general, in order to construct $\boldsymbol{\lambda}$ such that $\int_{0}^{\xi} [\psi(x)-\lambda(x)] \mathrm{d} x = \delta$ holds for any arbitrary $\delta>0$, a  maximum variable degree is required which may  tend to infinity, i.e., $d_v\to\infty$.
\end{remk}

\section{Numerical Results}\label{Numerical}
In this section, we evaluate the LDPC codes which are obtained by solution of the proposed optimization problems. For benchmark schemes, we consider the complexity-optimized code (COC) reported in \cite{XudongISIT} and performance-optimized codes (POCs) in \cite{Saeedi} and \cite{Shokrol2}. We consider both regular and irregular check degree distributions $\rho(x)=x^7$ and $\rho(x) = 0.5330x^6+0.4670x^7$, respectively, see footnote \footnote{This irregular check degree distribution is assumed in \cite{XudongISIT}. Therefore, to have a fair comparison, we also adopt this check degree distribution for the code design.}.  

\begin{figure}
\centering
\iftoggle{OneColumn}{%
\resizebox{0.8\linewidth}{!}
}{%
\resizebox{1\linewidth}{!}
}
{\psfragfig{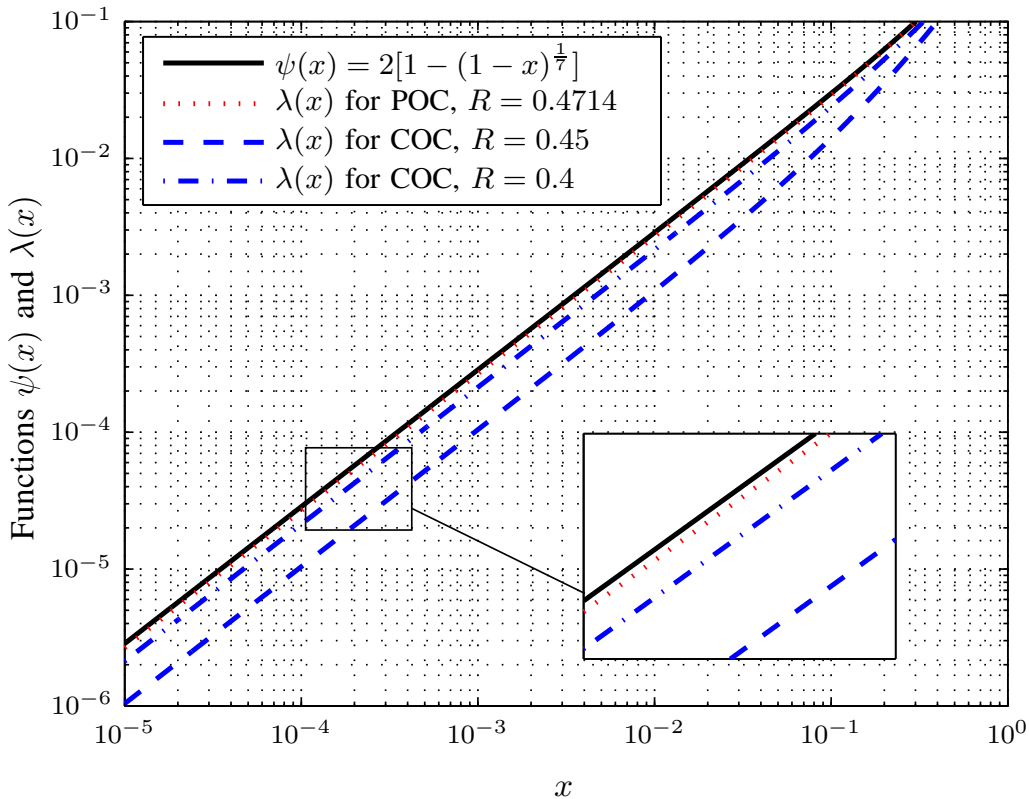}} 
    \caption{Modified density evolution: function $\lambda(x)$ for the performance-optimized and complexity-optimized codes obtained for $\rho(x)=x^7$, $d_v=16$, $\eta=10^{-5}$, and $\varepsilon=0.5$.}
    \label{DensCheck}
\end{figure}

Note that both, the proposed approximation of the number of decoding iterations given in (\ref{Approx}) and the utility function given in (\ref{Utility}), are based on the distance concept introduced for the modified density evolution in Section II-B. Therefore, the distance concept in Fig. \ref{DensCheck} is investigated for some performance-optimized and complexity-optimized codes. At first, we assume the following parameters for the code design $\rho(x)=x^7$, $d_v=16$, $\eta=10^{-5}$, and $\varepsilon=0.5$. Fig. \ref{DensCheck} shows the modified density evolutions introduced in Section II-B for the performance-optimized code obtained by means of (\ref{Rate1}) and complexity-optimized code obtained by means of (\ref{ProbApprox}). The maximum achievable code rate for the considered set of parameter is obtained as $R_{\max}=0.4714$ from  (\ref{Rate1}). We observe that the obtained variable degree distribution, $\lambda(x)$, for the performance-optimized code in  (\ref{Rate1}) is very close to function $\psi(x)=2[1-(1-x)^{\frac{1}{7}}]$ which leads to the high number of decoding iteration required to achieve the considered target residual erasure probability $\eta=10^{-5}$. However, if a lower code rate is considered, i.e., $R_d<R_{\max}$, we are able to design complexity-optimized codes which lead to a lower number of decoding iterations compared to performance-optimized  codes. Fig. \ref{DensCheck} shows that as the desired code rate decreases, the distance between $\lambda(x)$ designed in (\ref{ProbApprox}) and $\psi(x)$ increases which leads to a lower number of decoding iterations. 

\begin{figure}
\centering
\iftoggle{OneColumn}{%
\resizebox{0.8\linewidth}{!}
}{%
\resizebox{1\linewidth}{!}
}{
\psfragfig{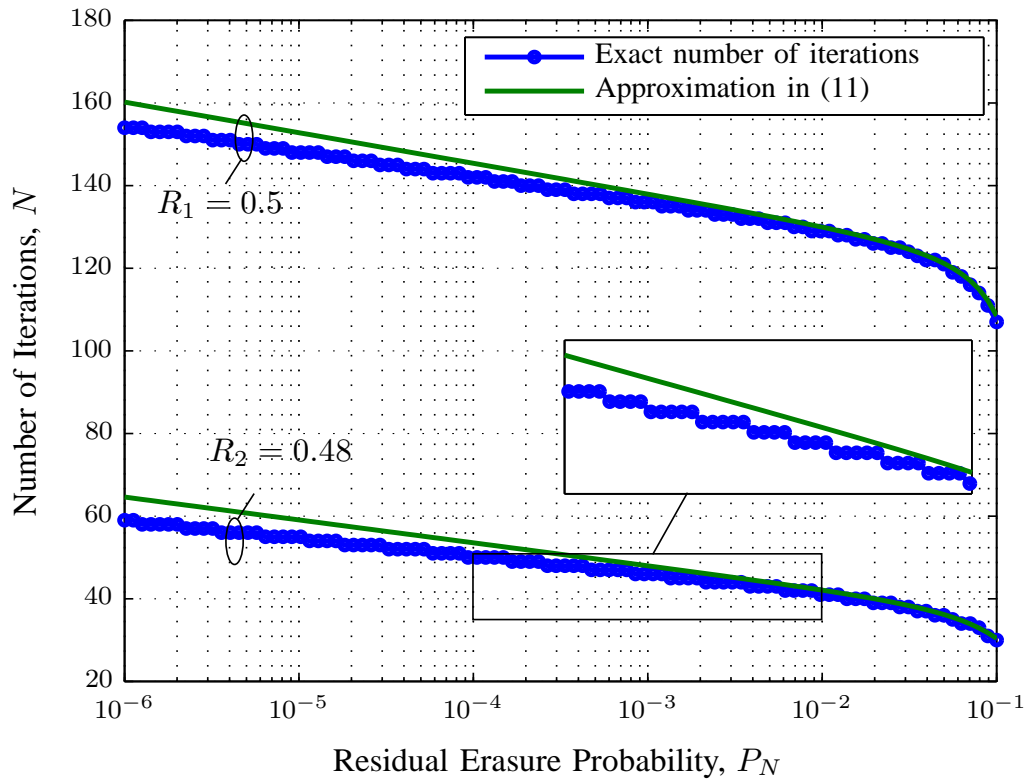}}  
    \caption{Number of iterations $N$ vs. the residual erasure probability $P_N$, $\rho(x) = 0.5330x^6+0.4670x^7$, $R_1=0.5$, $R_2=0.48$, and $\varepsilon=0.48$.}
    \label{Acc}
\end{figure}

Using the distance concept introduced for the modified density evolution, the approximation of the number of iterations (\ref{Approx}) is proposed.  In Fig. \ref{Acc},  the exact number of iterations and the proposed approximation of the number of iterations vs. the residual erasure probability are shown where the following parameters are used: $\rho(x) = 0.5330x^6 + 0.4670x^7$, $\lambda_1(x) =  0.2220x+0.3814x^2+0.1331x^8+0.2635x^{15}$, $\lambda_2(x) =  0.1881x + 0.4056x^2 + 0.0828x^8 + 0.3234x^{15}$, and $\varepsilon=0.48$. Note that $\lambda_1(x)$ and $\lambda_2(x)$ lead to code rates $R_1=0.48$ and $R_2=0.5$, respectively, where the channel capacity is $C=1-\varepsilon=0.52$. We observe that the approximation (\ref{Approx}) is quite accurate although we utilized the continuous assumption of the number of iterations in the derivation.  Moreover, changing the code rate does not noticeably change the accuracy of the proposed approximation. Furthermore, it can be easily seen from  Fig. \ref{Acc} that the function given in (\ref{NumberDef}) for the exact number of iterations is a non-differentiable function while the proposed approximation is a continuous function which significantly facilitates tackling the  optimization problem.

\begin{figure}
\centering
\iftoggle{OneColumn}{%
\resizebox{0.8\linewidth}{!}
}{%
\resizebox{1\linewidth}{!}
}{
\psfragfig{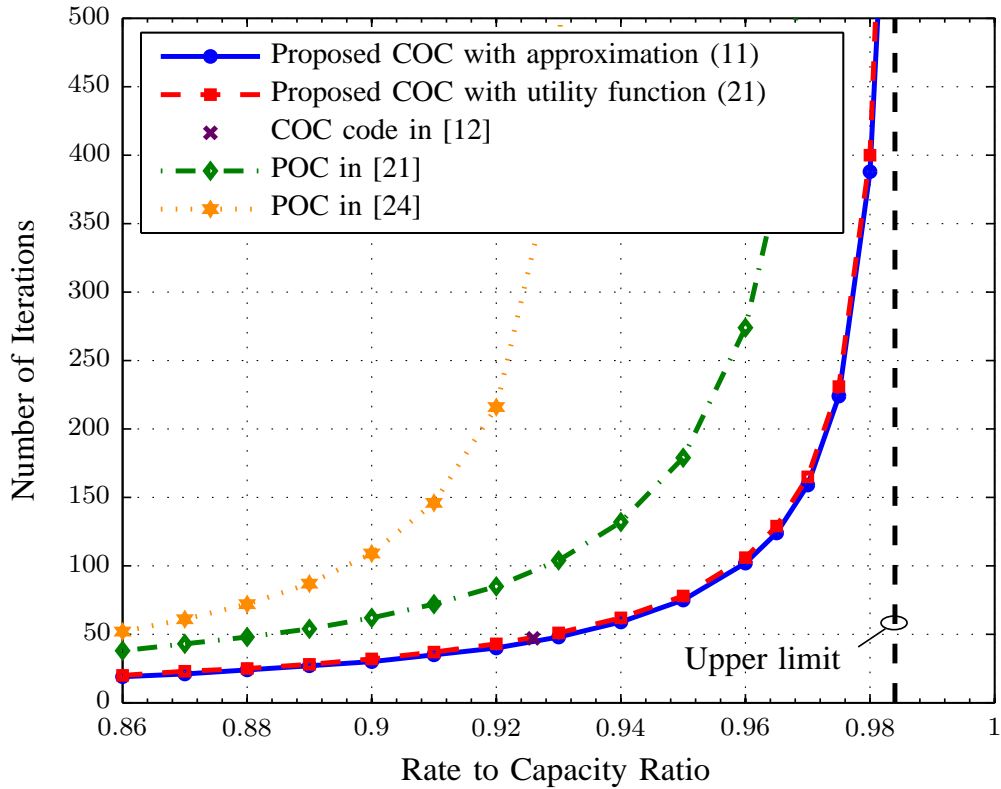}}    
    \caption{Number of iterations vs. rate to capacity ratio, $\frac{R}{1-\varepsilon}$,  for different codes, $\rho (x)= 0.5330x^6+0.4670x^7$, $d_v=16$, $R_d=0.5$, and $\eta=10^{-3}$. }
    \label{Comparison}
\end{figure}

In Fig. \ref{Comparison}, the number of decoding iterations vs. the rate to capacity ratio, i.e., $\frac{R}{1-\varepsilon}$, is depicted. We consider the following parameters for the code design $\rho (x)= 0.5330x^6 + 0.4670x^7$, $d_v=16$, $R=0.5$, and $\eta=10^{-3}$. The results for the proposed complexity-optimized codes designed by means of (\ref{ProbApprox}) (the approximation of the number of iterations) and (\ref{Op1}) (the utility function) are illustrated. We observe that the codes obtained with the proposed utility function leads to quite similar number of iterations compared to that obtained by the accurate approximation in (\ref{Approx}) which confirms the effectiveness of the proposed utility function. Note that the maximum achievable rate to capacity ratio for the considered set of parameter is obtained as $\frac{R}{1-\varepsilon}=0.984$ from the equivalent threshold maximization problem to the rate maximization problem in (\ref{Rate1}). Therefore, $\frac{R}{1-\varepsilon}=0.984$ is the upper limit for $d_v=16$ and the given $\rho(x)$. For any $\frac{R}{1-\varepsilon}>0.984$, both problems in (\ref{ProbApprox}) and (\ref{Op1}) become infeasible. As a performance benchmark, we probe the complexity-optimized code in \cite{XudongISIT} and performance-optimized codes in \cite{Saeedi} and \cite{Shokrol2}. Note that the codes proposed in \cite{XudongISIT} and \cite{Shokrol2} are obtained for a fixed rate $R_d=0.5$ which is the reason we consider the same rate. As can be seen from Fig. \ref{Comparison}, the proposed code in \cite{Saeedi}  requires a lower number of iterations compared to the code in \cite{Shokrol2}. However, both of them are outperformed by the new complexity-optimized codes in terms of the number of decoding iterations. Unfortunately, only one point is reported in \cite{XudongISIT} which coincides with the curves obtained via the proposed complexity-optimizing approach.

\begin{figure}
\centering
\iftoggle{OneColumn}{%
\resizebox{0.8\linewidth}{!}
}{%
\resizebox{1\linewidth}{!}
}{
\psfragfig{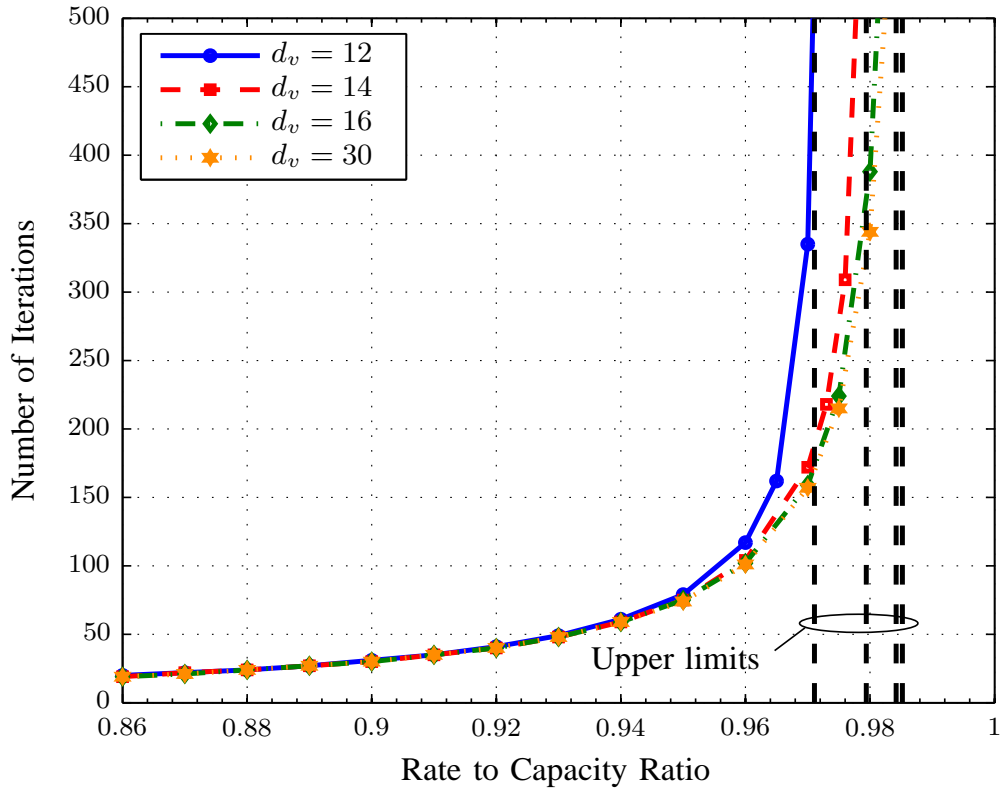}}   
    \caption{Number of Iterations vs. rate to capacity ratio, $\frac{R}{1-\varepsilon}$,  for different maximum variable node degree, $\rho(x)=0.5330 x^6+0.4670x^7$, $R=0.5$, and $\eta=10^{-3}$. }
    \label{DiffDv}
\end{figure}

\begin{figure}
\centering
\iftoggle{OneColumn}{%
\resizebox{0.8\linewidth}{!}
}{%
\resizebox{1\linewidth}{!}
}{
\psfragfig{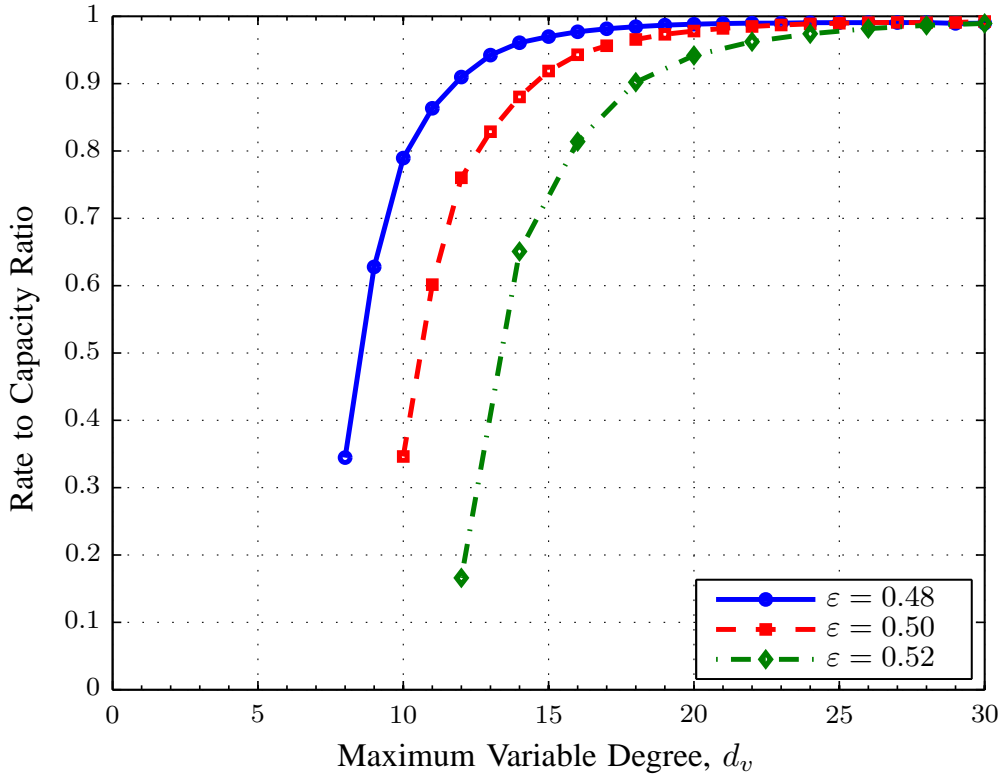}}     
    \caption{Maximum achievable rate to capacity ratio, $\frac{R}{1-\varepsilon}$, vs. maximum variable node degree, $d_v$ for different channel erasure probabilities and $\rho(x)=x^7$. }
    \label{RateDv}
\end{figure}

In Fig. \ref{DiffDv}, the effect of the value of the maximum variable degree $d_v$ on the proposed code design is investigated. Thereby, the same parameters are used as the ones in Fig. \ref{Comparison} and also the number of decoding iterations vs. the rate to capacity ratio is depicted. First, for a given rate to capacity ratio, the number of iterations decreases as $d_v$ increases. Second,  as $d_v$ increases the maximum achievable rate to capacity ratio, the upper limits in Fig. \ref{DiffDv},  increases. This is due to the fact that a higher $d_v$ leads to a larger  feasibility solution set in the optimization problems in (\ref{ProbApprox}), (\ref{Op1}) and (\ref{Rate1}) for the complexity-optimized and performance-optimized codes, respectively. However, the effect of increasing $d_v$ for low rate to capacity ratio is negligible. In order to illustrate the effect of increasing $d_v$ on the feasibility solution set for the considered optimization problem in (\ref{Op1}), we also plot the maximum rate to capacity ratio, $\frac{R}{1-\varepsilon}$, vs. maximum variable node degree, $d_v$, for $\rho(x)=x^7$ and different channel erasure probabilities $\varepsilon=0.48, 0.50, 0.52$, see Fig. \ref{RateDv}. Since as $d_v$ increases, the feasible set for the solution of the optimization problem in (\ref{Op1}) becomes larger which leads to a higher rate to capacity ratio. Moreover, as ultimately $d_v\to\infty$, we obtain $\frac{R}{1-\varepsilon}\to 1$. As an interesting observation here, at least for $d_v<20$, we observe from Fig. \ref{RateDv} that as $\varepsilon$ decreases, i.e., capacity increases, a higher rate to capacity is achieved. 

\begin{figure}
\centering
\iftoggle{OneColumn}{%
\resizebox{0.8\linewidth}{!}
}{%
\resizebox{1\linewidth}{!}
}{
\psfragfig{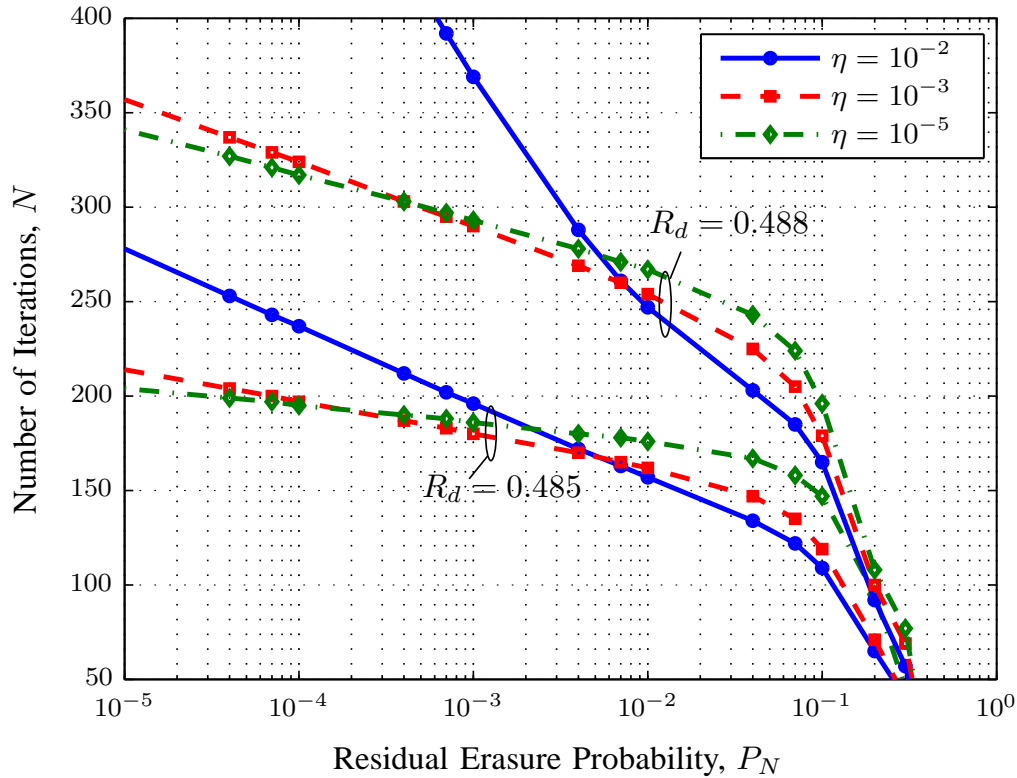}}     
    \caption{Number of Iterations vs. residual erasure probability for different values of design parameter $\rho(x)=0.5330 x^6+0.4670x^7$, $d_v=16$, $\varepsilon=0.5$, and $R_d=0.485,0.488$.}   
     \label{ErrorCom}
\end{figure}

Fig. \ref{ErrorCom} presents the number of decoding iterations vs. the residual erasure probability. We assume the following parameters for the code design $\rho(x)=0.5330 x^6+0.4670x^7$, $d_v=16$, $\varepsilon=0.5$, $R_d=0.485,0.488$, and $\eta=10^{-2},10^{-3},10^{-5}$. It can be seen that that each code requires lower number of iterations for the respective target residual erasure probability that is designed for. For instance, considering the set of parameters for $R_d=0.485$,  for the target residual erasure probability $10^{-5}$, the code that is designed for  $\eta=10^{-5}$ needs $204$ iterations while codes that designed for $\eta=10^{-3}$ and $\eta=10^{-2}$ need $214$ and $278$ number of iterations, respectively. Moreover, focusing on the result for  $\eta=10^{-5}$ as an example, it can be seen that in order to have a lower number of iterations for target erasure probability $10^{-5}$, first the code designed for $\eta=10^{-5}$ requires a higher number of iterations compared to the code designed for lower target residual erasure probabilities, i.e., $\eta=10^{-2},10^{-3}$, but finally, it outperforms them at the target erasure probability $\eta=10^{-5}$. This can be interpreted also in the modified density evolution in Fig. \ref{DensityEvol}. In particular, $\lambda(x)$ is designed for  $\eta=10^{-5}$ has a closer distance to $\psi(x)$ in the regimes $x>10^{-2}$ and $x>10^{-3}$ compared to that for $\lambda(x)$ that are designed for $\eta=10^{-2}$ and $\eta= 10^{-3}$, respectively, which leads to a higher number of iterations in these regimes. However, at this cost, the distance between $\lambda(x)$ which is designed for  $\eta=10^{-5}$ to $\psi(x)$ is higher in the regimes $x<10^{-2}$ and $x<10^{-3}$ compared to that for $\lambda(x)$ that are designed for $\eta=10^{-2}$ and $\eta= 10^{-3}$, respectively, which in total leads to a lower number of iterations at the target erasure probability $\eta=10^{-5}$.

Finally, in Table I, the degree distributions of the proposed codes used in this section and found by using CVX \cite{CVX} to solve the optimization problems (\ref{ProbApprox}), (\ref{Op1}) and (\ref{Rate1}) are presented. Note that all the presented coefficients of $\lambda(x)$ are rounded for four-digit accuracy. Optimized $\lambda(x)$ are given in Table I for different design criteria and parameters which allows some interpretations and intuitions. For instance, for the performance-optimized code in Fig. \ref{DensCheck} with code rate $R=0.4714$, we obtain $\lambda_2=0.2673$ while for the complexity-optimized codes with given code rates $R=0.45$ and $R=0.4$, the values are $\lambda_2=0.2126$ and $\lambda_2=0.1041$, respectively. Thus, we can conclude that, as a lower code rate is required,  we have to reduce the value of $\lambda_2$ for a complexity-optimized code. As an other example,  we compare the codes designed by  means of the proposed approximation (\ref{Approx}) with that obtained by means the proposed utility function (\ref{Utility}). The resulting $\lambda(x)$ corresponding to the graphs in Fig. \ref{Comparison} are similar  but not identical which roughly confirms the effectiveness of the proposed utility function. Last but not least, from the codes designed for different values of $d_v$ corresponding to Fig. \ref{DiffDv}, we can conclude that, for the rate-to-capacity ratio $\frac{R}{1-\varepsilon}=0.97$, increasing $d_v$ from $12$ to $16$, crucially changes the resulting complexity-optimized codes. Moreover, the maximum degree is non-zero for the codes with $d_v=12,16$. However, for $d_v=30$, we observe that the maximum degree with non-zero value is $19$ by which we can conclude that increasing $d_v$ more than $19$ cannot decrease the number of iterations for the considered rate-to-capacity ratio. This observation can also be confirmed from Fig. \ref{DiffDv} as for $d_v=12,16$ and $30$, the number of decoding iterations for   the rate-to-capacity ratio $\frac{R}{1-\varepsilon}=0.97$ are $335,159,$ and $157$, respectively.

\begin{table*}[!t]
\label{DegreeDist}
\caption{Several Examples of the Proposed Degree Distributions $(\boldsymbol{\lambda},\boldsymbol{\rho})$ Used in the Numerical Results.} 
\begin{center}
\iftoggle{OneColumn}{%
\scalebox{0.8}
}{%
\scalebox{0.9}
}{
\begin{tabular}{|| c | c | c ||}
  \hline                  
$\lambda(x)$ & $\rho(x)$ & \text{Comments} \\ \hline
$0.2673x+0.2107x^2+0.5220x^{15}$ & $x^7$ & \text{Fig. \ref{DensCheck}, POC,} $R=0.4714$\\ \hline
$0.2126x+0.2650x^2+0.5224x^{15}$ & $x^7$ & \text{Fig. \ref{DensCheck}, COC,} $R=0.4500$\\ \hline
$0.1041x+0.3704x^2+0.5255x^{15}$ & $x^7$ & \text{Fig. \ref{DensCheck}, COC,} $R=0.4000$\\ \hline  \hline  
$0.1881x+0.4056x^2+0.0828x^8+0.3234x^{15}$ & $0.5330x^6+0.4670x^7$ & \text{Fig. \ref{Acc}, } $R=0.50$\\ \hline  
$0.2220x+0.3814x^2+0.1331x^8+0.2635x^{15}$ &$0.5330x^6+0.4670x^7$ & \text{Fig. \ref{Acc}, } $R=0.45$\\ \hline \hline
$0.1555x+0.4993x^2+0.0348x^{12}+0.3104x^{13}$ &$0.5330x^6+0.4670x^7$ & \text{Fig. \ref{Comparison}, Prop. Code with Approx.,} $\frac{R}{1-\varepsilon}=0.90$\\ \hline 
$0.1301x+0.5279x^2+0.2651x^{11}+0.0769x^{12}$ &$0.5330x^6+0.4670x^7$ & \text{Fig. \ref{Comparison}, Prop. Code with Utility,} $\frac{R}{1-\varepsilon}=0.90$\\ \hline
$0.2172x+0.3888x^2+0.0470x^{9}+0.1559x^{10}+0.1911x^{15}$ &$0.5330x^6+0.4670x^7$ & \text{Fig. \ref{Comparison}, Prop. Code with Approx.,} $\frac{R}{1-\varepsilon}=0.94$\\ \hline 
$0.2056x+0.3971x^2+0.1859x^8+0.2114x^{15}$ &$0.5330x^6+0.4670x^7$& \text{Fig. \ref{Comparison}, Prop. Code with Utility,} $\frac{R}{1-\varepsilon}=0.94$\\ \hline
$0.2986x+0.1833x^2+0.1602x^4+0.0406x^5+0.3173x^{15}$ &$0.5330x^6+0.4670x^7$ & \text{Fig. \ref{Comparison}, Prop. Code with Approx.,} $\frac{R}{1-\varepsilon}=0.98$\\ \hline 
$0.2940x+0.2016x^2+0.1013x^4+0.0901x^5+0.3130x^{15}$ &$0.5330x^6+0.4670x^7$& \text{Fig. \ref{Comparison}, Prop. Code with Utility,} $\frac{R}{1-\varepsilon}=0.98$\\ \hline \hline
$0.2793x+0.2774x^2+0.4434x^{12}$ &$0.5330x^6+0.4670x^7$& \text{Fig. \ref{DiffDv},} $\frac{R}{1-\varepsilon}=0.97$, $d_v=12$\\ \hline
$0.2743x+0.2571x^2+0.1484x^5+0.0181x^6+0.3022x^{15}$ &$0.5330x^6+0.4670x^7$& \text{Fig. \ref{DiffDv},} $\frac{R}{1-\varepsilon}=0.97$, $d_v=16$\\ \hline
$0.2715x+0.2721x^2+0.0941x^6+0.1390x^7+0.2233x^{18}$ &$0.5330x^6+0.4670x^7$& \text{Fig. \ref{DiffDv},} $\frac{R}{1-\varepsilon}=0.97$, $d_v=30$\\ \hline \hline
$0.2977x+0.0949x^2+0.1901x^3+0.4173x^{15}$ &$x^7$& \text{Fig. \ref{RateDv},} $d_v=16$, $\varepsilon=0.48$\\ \hline
$0.2673x+0.2107x^2+0.5220x^{15}$ &$x^7$& \text{Fig. \ref{RateDv},} $d_v=16$, $\varepsilon=0.50$\\ \hline
$0.2749x+0.0828x^2+0.6424x^{15}$ &$x^7$& \text{Fig. \ref{RateDv},} $d_v=16$, $\varepsilon=0.52$\\ \hline \hline
$0.3051x+0.0589x^2+0.2709x^3+0.3651x^{15}$ &$0.5330x^6+0.4670x^7$& \text{Fig. \ref{ErrorCom},} $R=0.488$, $\eta=10^{-2}$\\ \hline
$0.2888x+0.1681x^2+0.0628x^3+0.1206x^4+0.3598x^{15}$ &$0.5330x^6+0.4670x^7$& \text{Fig. \ref{ErrorCom},} $R=0.488$, $\eta=10^{-3}$\\ \hline
$0.2811x+0.2077x^2+0.1528x^4+0.3584x^{15}$ &$0.5330x^6+0.4670x^7$& \text{Fig. \ref{ErrorCom},} $R=0.488$, $\eta=10^{-5}$\\ \hline
\end{tabular}
}
\end{center}
\vspace{-0.5cm}
\end{table*}

\section{Conclusion}
The design of complexity-optimized LDPC codes for the BEC is considered in this paper. To this end, a quite accurate approximation of the number of iterations is proposed based on the concept of  performance-complexity tradeoff. For any given check degree distribution, a complexity-optimizing problem is formulated to find a variable degree distribution for a given finite maximum variable node degree such that the number of required decoding iterations to achieve a certain target residual erasure probability is minimized and a given code rate is guaranteed. It is shown that the optimization problem is convex in the optimization variables, however, it belongs to the class of semi-infinite programming, i.e., it has infinite number of constraints. To facilitate obtaining a complexity-optimized code, first a lower bound on the number of iterations is proposed. Then,  a utility function corresponding to the number of iterations is introduced based on this proposed lower bound. It is shown that the resulting optimization problem has a SDP representation and thus
can be solved much more efficiently compared to semi-infinite problems. Numerical results confirmed the effectiveness of the proposed methods for code design.

\appendices


\section{Proof of Theorem \ref{SDP}}
\label{AppSDP}

In this appendix, the SDP representability of constraint $\mathrm{C1}$ in (\ref{Op2}) is shown. To this end, we first rewrite $\psi(x)$ in a polynomial form $\psi'(x) = {\underset{M\rightarrow \infty}{\textrm{lim}}} \sum\nolimits_{i = 2}^M {(i-1)T_i x^{i - 2} }$  where $T_i$ are the Taylor expansion coefficients of $\psi(x)$ around $x=0$. Let
\begin{equation}
   f(x) = \psi(x) - \lambda(x) = \sum\limits_{i = 1}^{D} {f_i x^{i} }
\end{equation}
where $D=\max\{d_v,M\}$. Assuming a relatively large value of $M$ to guarantee the accuracy of the approximation $\psi(x) \approx  \sum\nolimits_{i = 2}^M {T_i x^{i - 1} }$ \cite{Saeedi}, we consider the case $M > d_v$ for the sake of simplicity of presentation. Thus, we obtain $D = M-1$ and $f_i$ is given by
\begin{IEEEeqnarray}{ll}
   f_i =
   \left\{ \begin{array}{ll}
   T_{i+1} - \lambda_{i+1},  & i = 1,\dots, d_v - 1, \\
   T_{i+1}, & \mathrm{ otherwise }
   \end{array} \right .
 \nonumber
\end{IEEEeqnarray}
Note that affine mapping preserves the semi-definite representability of a set\cite{Nemirovski}. Here, we use the following mapping 
\begin{IEEEeqnarray}{ll}\label{Mapping}
x=(\xi-\tilde{\zeta})\frac{y^2}{1+y^2}+\tilde{\zeta},
\end{IEEEeqnarray}
which maps $x\in[\tilde{\zeta},\,\,\xi]$ to $y\in(-\infty, \,\,+\infty)$. Note that the mapping (\ref{Mapping}) has to be affine in the optimization variables $\boldsymbol{\lambda},t$ not in $x$. Using this mapping, constraint $\mathrm{C1}$ in (\ref{Op2}) is rewritten as follows
\iftoggle{OneColumn}{%
\begin{IEEEeqnarray}{ll}\label{MappResult}
  f(x)> t\psi'(x), \quad &\mathrm{for} \,\, x \in [\tilde{\zeta},\,\,\xi] \,\, \Leftrightarrow \nonumber \\
    &f\hspace{-0.1cm}\left(\hspace{-0.1cm}(\xi\Minus\tilde{\zeta})\hspace{-0.1cm}\left(\hspace{-0.1cm}\frac{y^2}{1+y^2}\hspace{-0.1cm}\right)\Add\tilde{\zeta}\hspace{-0.1cm}\right)\hspace{-0.1cm} \Great t \psi'\hspace{-0.1cm}\left(\hspace{-0.1cm}(\xi\Minus\tilde{\zeta})\hspace{-0.1cm}\left(\hspace{-0.1cm}\frac{y^2}{1+y^2}\hspace{-0.1cm}\right)\Add \tilde{\zeta} \hspace{-0.1cm}\right)\hspace{-0.1cm},  \, \mathrm{for} \, y\hspace{-0.1cm} \in\hspace{-0.1cm} \mathbb{R} \qquad
\end{IEEEeqnarray}
}{%
\begin{align}\label{MappResult}
  f(x)> t\psi'(x), \quad \mathrm{for} \,\, x \in [\tilde{\zeta},\,\,\xi] \Leftrightarrow \qquad\qquad\qquad\qquad\qquad\quad\,\,\,\nonumber \\
    f\hspace{-0.1cm}\left(\hspace{-0.1cm}(\xi\Minus\tilde{\zeta})\hspace{-0.1cm}\left(\hspace{-0.1cm}\frac{y^2}{1+y^2}\hspace{-0.1cm}\right)\Add\tilde{\zeta}\hspace{-0.05cm}\right)\hspace{-0.05cm} \Great t \psi'\hspace{-0.1cm}\left(\hspace{-0.1cm}(\xi\Minus\tilde{\zeta})\hspace{-0.1cm}\left(\hspace{-0.1cm}\frac{y^2}{1+y^2}\hspace{-0.1cm}\right)\Add \tilde{\zeta} \hspace{-0.05cm}\right)\hspace{-0.1cm},  \, \mathrm{for} \, y\hspace{-0.05cm} \in\hspace{-0.05cm} \mathbb{R} \quad
\end{align}
}
By multiplying both side of (\ref{MappResult}) by positive value  $(1+y^2)^D$, we obtain
\iftoggle{OneColumn}{%
\begin{IEEEeqnarray}{llll}\label{DenNom}
 (1+y^2)^{D} f\left(\frac{\xi y^2+\tilde{\zeta}}{1+y^2}\right) &= (1+y^2)^{D} \left[ \sum\limits_{i = 1}^{D} {f_i \left(\frac{\xi y^2+\tilde{\zeta}}{1+y^2}\right)^{i} } \right] & \nonumber \\
  & >  t (1+y^2)^{D}\left[\sum\limits_{i = 1}^D {i T_{i+1} \left(\frac{\xi y^2+\tilde{\zeta}}{1+y^2}\right)^{i - 1} }\right]\hspace{-0.1cm}, \,\, \mathrm{for}\,\,y \in \mathbb{R}. \qquad &
\end{IEEEeqnarray}
}{%
\begin{IEEEeqnarray}{llll}\label{DenNom}
 (1+y^2)^{D} f\left(\frac{\xi y^2+\tilde{\zeta}}{1+y^2}\right) = (1+y^2)^{D} \left[ \sum\limits_{i = 1}^{D} {f_i \left(\frac{\xi y^2+\tilde{\zeta}}{1+y^2}\right)^{i} } \right] & \nonumber \\
  \quad >  t (1+y^2)^{D}\left[\sum\limits_{i = 1}^D {i T_{i+1} \left(\frac{\xi y^2+\tilde{\zeta}}{1+y^2}\right)^{i - 1} }\right]\hspace{-0.1cm}, \,\, \mathrm{for}\,\,y \in \mathbb{R}. \qquad &
\end{IEEEeqnarray}
}
Removing the identical terms in the nominators and the respective denominators and moving all terms in one side of the inequality lead to
\iftoggle{OneColumn}{%
\begin{IEEEeqnarray}{ll}\label{Equ16}
   \sum\limits_{i = 1}^{D} f_i {\tilde{\zeta}}^i \left(\frac{\xi}{\tilde{\zeta}} y^2+1\right)^{i} \left(y^2+1\right)^{D - i}  
   -\hspace{-0.1cm} \sum\limits_{i = 1}^D \hspace{-0.1cm} {i T_{i+1} {\tilde{\zeta}}^{i-1} \hspace{-0.1cm} \left(\hspace{-0.1cm}\frac{\xi}{\tilde{\zeta}} y^2\Add 1\hspace{-0.1cm}\right)^{i - 1}\hspace{-0.4cm} (y^2\Add 1)^{D-i+1}} t \Great 0\hspace{-0.03cm}, \,  \mathrm{for}\, y \hspace{-0.1cm}\in \hspace{-0.1cm}\mathbb{R}
\end{IEEEeqnarray}
}{%
\begin{IEEEeqnarray}{ll}\label{Equ16}
   \sum\limits_{i = 1}^{D} f_i {\tilde{\zeta}}^i \left(\frac{\xi}{\tilde{\zeta}} y^2+1\right)^{i} \left(y^2+1\right)^{D - i}   \nonumber \\
   -\hspace{-0.1cm} \sum\limits_{i = 1}^D \hspace{-0.1cm} {i T_{i+1} {\tilde{\zeta}}^{i-1} \hspace{-0.1cm} \left(\hspace{-0.1cm}\frac{\xi}{\tilde{\zeta}} y^2\Add 1\hspace{-0.1cm}\right)^{i - 1}\hspace{-0.4cm} (y^2\Add 1)^{D-i+1}} t \Great 0\hspace{-0.03cm}, \,  \mathrm{for}\, y \hspace{-0.1cm}\in \hspace{-0.1cm}\mathbb{R} \quad
\end{IEEEeqnarray}
}

Next, let the left hand side of (\ref{Equ16}) be written as $\pi(y) = \sum_{k = 0}^{D}{\pi_k y^{2k}}$. Then, by considering the binomial expansion, $(1 + x)^n = \sum\limits_{j = 0}^n{ {n \choose j} x^j }$ with ${n \choose j}=\frac{n!}{j!(n-j)!}$, $\pi(y)$ is expanded by 
\iftoggle{OneColumn}{%
\begin{IEEEeqnarray}{lll}\label{f1_2}
 \pi(y) = \sum\limits_{k = 0}^{D}{\pi_k y^{2k}} & = \sum\limits_{i = 1}^{D} f_i {\tilde{\zeta}}^i  \left( \sum\limits_{j = 0}^{i}{a_j y^{2j}}\right) \left( \sum\limits_{j = 0}^{D - i}{b_j y^{2j}} \right)  \nonumber \\ 
 &- \hspace{-0.1cm}\sum\limits_{i = 1}^D {i T_{i+1}{\tilde{\zeta}}^{i-1} \hspace{-0.1cm}\left(\hspace{-0.05cm} \sum\limits_{j = 0}^{i-1}{c_j y^{2j}}\hspace{-0.1cm}\right)\hspace{-0.1cm} \left(\hspace{-0.1cm} \sum\limits_{j = 0}^{D-i+1}\hspace{-0.1cm}{d_j y^{2j}}\hspace{-0.1cm}\right) \hspace{-0.1cm}t} \Great 0\hspace{-0.03cm}, \,\mathrm{for} \, y \hspace{-0.1cm}\in\hspace{-0.1cm} \mathbb{R} \qquad
\end{IEEEeqnarray}
}{%
\begin{IEEEeqnarray}{lll}\label{f1_2}
 \pi(y) = \sum\limits_{k = 0}^{D}{\pi_k y^{2k}} = \sum\limits_{i = 1}^{D} f_i {\tilde{\zeta}}^i  \left( \sum\limits_{j = 0}^{i}{a_j y^{2j}}\right) \left( \sum\limits_{j = 0}^{D - i}{b_j y^{2j}} \right)  \nonumber \\ 
 - \hspace{-0.1cm}\sum\limits_{i = 1}^D {i T_{i+1}{\tilde{\zeta}}^{i-1} \hspace{-0.1cm}\left(\hspace{-0.05cm} \sum\limits_{j = 0}^{i-1}{c_j y^{2j}}\hspace{-0.1cm}\right)\hspace{-0.1cm} \left(\hspace{-0.1cm} \sum\limits_{j = 0}^{D-i+1}\hspace{-0.1cm}{d_j y^{2j}}\hspace{-0.1cm}\right) \hspace{-0.1cm}t} \Great 0\hspace{-0.03cm}, \,\mathrm{for} \, y \hspace{-0.1cm}\in\hspace{-0.1cm} \mathbb{R} \qquad
\end{IEEEeqnarray}
}
where $a_j,b_j,c_j$, and $d_j$ are given in (\ref{Coefficients}).
In order to explicitly obtain the coefficients $\pi_k,\,\,k=0,\dots,D$ in terms of the known parameters, the polynomial multiplication rule (convolution rule)
\begin{IEEEeqnarray}{l}
\left(\hspace{-0.05cm} \sum\limits_{i = 0}^{n}{\alpha_{i}x^{i}}\hspace{-0.05cm} \right)\hspace{-0.1cm}\left(\hspace{-0.05cm} \sum\limits_{i = 0}^{m}{\beta_{i}x^{i}}\hspace{-0.05cm} \right)
\Equal\sum\limits_{i = 0}^{n+m}\hspace{-0.05cm} {\gamma_{i}x^{i}},\,\, \text{with}\,\, \gamma_i\Equal\sum\limits_{j = 0}^{i}\hspace{-0.05cm} {\alpha_{i-j}\beta_{j}} \quad
\end{IEEEeqnarray}
is applied.  Using the above rule, the
coefficients $\pi_k,\,\,k=0,\dots,D$ are obtained as $\pi_0\Equal0$ and
\iftoggle{OneColumn}{%
\begin{IEEEeqnarray}{ll}\label{f1_i}
\pi_k = \sum\limits_{i = 1}^{D}{ f_i {\tilde{\zeta}}^i \left( \sum\limits_{j = 0}^{k}{a_{k - j}b_j}\right)} 
 - \sum\limits_{i = 1}^{D}{ iT_i {\tilde{\zeta}}^{i-1} \left( \sum\limits_{j = 0}^{k}{c_{k - j}d_j}\right)} t,\quad k=1,\dots, D
\end{IEEEeqnarray}
}{%
\begin{IEEEeqnarray}{ll}\label{f1_i}
\pi_k &= \sum\limits_{i = 1}^{D}{ f_i {\tilde{\zeta}}^i \left( \sum\limits_{j = 0}^{k}{a_{k - j}b_j}\right)} \nonumber \\
&\quad - \sum\limits_{i = 1}^{D}{ iT_i {\tilde{\zeta}}^{i-1} \left( \sum\limits_{j = 0}^{k}{c_{k - j}d_j}\right)} t,\quad k=1,\dots, D
\end{IEEEeqnarray}
}

Now, we are ready to use Lemma \ref{Nemirov} to show the SDP representation of constraint $\mathrm{C1}$ in (\ref{Op2}). In particular, the interval in constraint $\mathrm{C1}$ in (\ref{Op2}), i.e., $[\tilde{\zeta},\,\, \xi]$, is mapped to $\left(-\infty, +\infty \right)$ where all coefficients $\pi_k$ are in affine form of optimization variables $\left[ \boldsymbol{\lambda},t \right]$. Therefore, by using Lemma \ref{Nemirov}, constraint $\mathrm{C1}$ in (\ref{Op2}) has the SDP representation as introduce in Theorem \ref{SDP}. This completes the proof.

\newpage
\bibliographystyle{IEEEtran}
\bibliography{Ref_11_01_2014}

\end{document}